\documentclass[aps,showpacs,superscriptaddress,nofootinbib,preprintnumbers,showkeys]{revtex4-1}
\usepackage{epsfig,amssymb,amsmath,psfrag,epstopdf,color}
\pdfoutput=1
\usepackage{graphicx}
\usepackage{multirow}
\usepackage{ulem}
\usepackage[breaklinks,bookmarksopen=true]{hyperref}
\usepackage{natbib}
\definecolor{linkcolor}{rgb}{0.0,0.3,0.5}
\usepackage[all]{hypcap}
\usepackage[T1]{fontenc}
\usepackage[utf8]{inputenc}
\usepackage{slashed}
\usepackage{booktabs}
\usepackage{lipsum}
\usepackage[table]{xcolor}
\usepackage{comment}
\usepackage{adjustbox,array}
\usepackage[section]{placeins} 
\usepackage{hyperref}
\usepackage{stmaryrd}

\definecolor{rosso}{cmyk}{0,1,1,0.4}
\definecolor{rossos}{cmyk}{0,1,1,0.55}
\definecolor{rossoc}{cmyk}{0,1,1,0.2}
\definecolor{blu}{cmyk}{1,1,0,0.3}
\definecolor{blus}{cmyk}{1,1,0,0.6}
\definecolor{bluc}{cmyk}{1,1,0,0.1}
\definecolor{verde}{cmyk}{0.92,0,0.59,0.25}
\definecolor{verdec}{cmyk}{0.92,0,0.59,0.15}
\definecolor{verdes}{cmyk}{0.92,0,0.59,0.4}

\DeclareUnicodeCharacter{2212}{\ensuremath{-}}

\hypersetup{
 colorlinks=true,
 linkcolor=verdec,
 urlcolor=verdec,
 citecolor=verdec
}

\begin{document}

\preprint{CTPU-PTC-23-37}

\title{Boosting dark matter searches at muon colliders with Machine Learning: the mono-Higgs channel as a case study}

\def\andname{\hspace*{-0.5em}} 
\author{\textsc{Mohamed Belfkir}}
\email{m_belfkir@uaeu.ac.ae}
\affiliation{Department of physics, United Arab Emirates University, Al-Ain, UAE}
\author{\textsc{Adil Jueid}}
\email{adiljueid@ibs.re.kr}
\affiliation{Particle Theory and Cosmology Group, Center for Theoretical Physics of the Universe, Institute for Basic Science (IBS), 34126 Daejeon, Republic of Korea}
\author{and \textsc{Salah Nasri}}
\email{snasri@uaeu.ac.ae; salah.nasri@cern.ch}
\affiliation{Department of physics, United Arab Emirates University, Al-Ain, UAE}
\affiliation{The Abdus Salam International Centre for Theoretical Physics, Strada Costiera 11, I-34014, Trieste, Italy}

\begin{abstract}
The search for dark-matter (DM) candidates at high-energy colliders is one of the most promising avenues to understand the nature of this elusive component of the universe. Several searches at the Large Hadron Collider (LHC) have strongly constrained a wide range of simplified models. The combination of the bounds from the LHC with direct-detection experiments exclude the most minimal scalar singlet DM model. To address this, Lepton portal DM models are suitable candidates where DM is predominantly produced at lepton colliders since the DM candidate only interacts with the lepton sector through a mediator that carries a lepton number.  In this work, we analyse the production of DM pairs in association with a Higgs boson decaying into two bottom quarks at future muon colliders in the framework of the minimal lepton portal DM model. It is found that the usual cut-based analysis methods fail to probe heavy DM masses for both the resolved (where the decay products of the Higgs boson can be resolved as two well-separated small-$R$ jets) and the merged (where the Higgs boson is clustered as one large-$R$ jet). We have then built a search strategy based on Boosted-Decision Trees (BDTs). We have optimised the hyperparameters of the BDT model to both have a high signal-to-background ratio and to avoid overtraining effects. We have found very important enhancements of the signal significance with respect to the cut-based analysis by factors of $8$--$50$ depending on the regime (resolved or merged) and the benchmark points. Using this BDT model on a one-dimensional parameter space scan we found that future muon colliders with $\sqrt{s}=3$ TeV and ${\cal L} = 1~{\rm ab}^{-1}$ can exclude DM masses up to $1$ TeV at the $95\%$ CL.
\end{abstract}

\maketitle


\newcommand{\ssection}[1]{{\em #1.\ }}
\newcommand{\rem}[1]{\textbf{#1}}

\section{Introduction}
\label{sec:Intro}

Weakly interacting massive particles (WIMPs) are suitable candidates which extend the Standard Model (SM) in order to solve the dark-matter (DM) problem \cite{Jungman:1995df,Bergstrom:2000pn,Bertone:2004pz,Feng:2010gw}. In particular WIMPs of mass of about $100$ GeV yield a DM density of $\Omega_{\rm DM} h^2 = 0.1198\pm 0.0015$ in agreement with the \textsc{Planck} observation \cite{Ade:2015xua}. On the other hand, WIMPs can be connected to other problems like the hierarchy problem, baryon asymmetry in the universe or the nature of neutrino mass generation mechanism. The understanding of the nature of WIMPs has driven many searches in direct detection, indirect detection and collider experiments. At the Large Hadron Collider (LHC), the ATLAS and CMS collaborations have carried several searches in channels dubbed as mono--X, {\it i.e.} channels where a visible particle recoils against a large transverse missing energy ($E_{T}^{\rm miss}$). Unfortunately all these searches were unsuccessful to find any interesting signal beyond the expected SM backgrounds and model-independent bounds were put on the DM mass versus the production cross section. On the other hand, strong bounds from direct detection experiments were also imposed as a consequence of the absence of new DM signals \cite{Aprile:2018dbl, Cui:2017nnn}. By combining the results from both  collider and direct detection experiments, we can narrow down the range of plausible beyond the SM scenarios aimed at addressing the DM problem. While the constraints from direct-detection experiments can be circumvented if the DM candidate is a right-handed fermion for example\footnote{Studies have shown that right-handed fermion singlets interact with the nucleons solely via loops and therefore their interaction rates are small thanks to loop suppression factors even for couplings of order one (see {\it e.g.} Refs. \cite{Okada:2013rha,Mohan:2019zrk,Belanger:2021smw}).},  earlier and current collider searches strictly constrain minimal models (see for example a recent reinterpretation of multijet+$E_{T}^{\rm miss}$ searches \cite{Arina:2023msd}). To avoid these issues, there is a possibility that the DM candidate couples {\it solely} to the lepton sector of the SM. In such scenarios the DM is called leptophilic and the models that contain such a DM are called lepton portal models. In the case of fermionic DM candidates, the interaction Lagrangian resembles to that of slepton--lepton--neutralino interaction in supersymmetric models. The most minimal choice in this case would consist of extending the SM with only two $SU(2)_L$ singlets. The phenomenology of these minimal models has been studied in Refs. \cite{Liu:2013gba,Bai:2014osa,Chang:2014tea,Garny:2015wea,Jueid:2020yfj,Horigome:2021qof,Liu:2021mhn,Jueid:2023abcs}. \\

The LHC programme for DM searches is still not over with the possibility to probe DM masses as high as $0.5$--$2$ TeV depending on the model. However, there is {\it a priori} no reason to not consider alternative future collider experiments which can both achieve high center-of-mass energies and provide very clean environments. Muon colliders are expected to provide both these characterestics at relatively low costs as compared to {\it e.g.} future circular hadron colliders (FCC--hh). There is a growing interests on the physics potential of muon colliders since they can probe new physics beyond the SM at very high scales \cite{Delahaye:2019omf,Long:2020wfp,AlAli:2021let}. There are many reasons for this interest. First, muon colliders can achieve small signal-to-background ratios as compared to the LHC. Second, given that muons are elementary particles, the center-of-mass energy required to achieve the same beam-level cross section is always orders of magnitude smaller than that in $pp$ collisions. Finally, at energies much higher than the production threshold of heavy resonances, muon colliders become vector-boson colliders where the dominant production channels occur through vector-boson fusion (VBF) \cite{Costantini:2020stv,Ruiz:2021tdt} in which case many processes in beyond the SM are free of backgrounds. Extensive studies of both the SM and beyond the SM have been carried in the literature \cite{Capdevilla:2020qel,Chiesa:2020awd,Han:2020uid,Han:2020uak,Yin:2020afe,Huang:2021nkl,Capdevilla:2021rwo,Capdevilla:2021fmj,Asadi:2021gah,Casarsa:2021rud,Liu:2021akf,Han:2021udl,Han:2021kes,Han:2021lnp,Lv:2022pts,Liu:2022byu,Azatov:2022itm,Yang:2022fhw,Bao:2022onq,Chen:2022msz,Homiller:2022iax,Jueid:2023abcs,Huang:2021biu,Sun:2023cuf,Zhang:2023yfg,Yang:2023ojm,Garosi:2023bvq,Dong:2023nir,Fridell:2023gjx,Garosi:2023bvq,Chowdhury:2023imd,Belyaev:2023yym,Yang:2023gos,Belfkir:2023lot,Forslund:2023reu,Liu:2023yrb,Jana:2023ogd,Dasgupta:2023zrh,Ghosh:2023xbj,Ouazghour:2023plc,Frixione:2023gmf}.  \\ 
  
The main production mechanism of DM at muon colliders is through mono--$X$ channels. For the case of muon colliders, the phenomenology of DM in mono-photon, and mono-muon channels has been performed at the parton level in Ref. \cite{Han:2020uak}. The mono-Higgs channel is however very unique since the SM Higgs boson is the only particle that can be produced from the interaction with the dark sector due to the smallness of the muon-Yukawa coupling unlike the mono-photon channel which can also be produced from Initial-State Radiation (ISR). Therefore, the mono-Higgs channel can be a very important channel to study the characteristics of the underlying model. The mono-Higgs channel has been suggested sometime ago in Ref. \cite{Carpenter:2013xra} and studied extensively for both hadron colliders and lepton colliders (see {\it e.g.} Refs. \cite{Berlin:2014cfa,No:2015xqa,Abdallah:2016vcn,Ghorbani:2016edw,Basso:2015aee,Antusch:2015gjw,Petrov:2013nia,Baum:2017gbj,Ahriche:2018ger,Argyropoulos:2021sav,Hammad:2022lzo,Banerjee:2021hfo,Bhowmik:2020spw}). In Ref. \cite{Jueid:2023abcs} a comprehensive analysis of the DM production within minimal lepton portal DM model has been done including the mono-Higgs production. This article presents a first study in a serie of upcoming works where we fully analyse the most prominent channels for DM at muon colliders and using state-of-art tools. In this study, we analyse the production of DM in association with a Higgs boson decaying into bottom quarks at future muon colliders using center-of-mass energy of $1$ TeV and a total luminosity of $1$ ab$^{-1}$. The final state consists of jets and missing energy. We use both the resolved regime where the decay products of the Higgs boson are two well-separated jets and the merged regime where the Higgs boson is identified with a single large-$R$ jet. First we use cut-based analysis inspired from the searches of DM in the mono-Higgs channel that have been carried by the ATLAS and CMS collaborations \cite{ATLAS:2021shl,CMS:2018zjv}. It is found that cut-based search strategies are not sensitive to heavy DM masses. We then employ a boosted-decision Tree (BDT) algorithm as implemented in \textsc{XGBoost} to further improve the signal-to-background ratio. We found that the BDT model leads to very important enhancements of the significance by large factors of $8$--$50$ depending on the analysis region and the DM mass. We find that DM masses up to $1$ TeV can be probed at future muon colliders if one employs BDTs. \\

The rest of this paper is organised as follows. In section \ref{sec:model} we briefly describe the theoretical setup including the benchmark points and the scanning procedure. We define the technical setup along with the discussion about the signal and the background cross sections and the object definitions at the detector level in section \ref{sec:technics}. Section \ref{sec:cut} is devoted to a detailed cut-based analysis. We then study the sensitivty reach using the BDT model in section \ref{sec:BDTs}. We conclude in section \ref{sec:conclusions}. \\

\section{Theoretical setup}
\label{sec:model}
In this section, we discuss the theoretical setup used in this work. We briefly introduce the model, its particle content and its parameters.  We close this section by a discussion of the benchmark points and the scanning procedure to be used in the collider analysis.

\subsection{The model}

We extend the SM by two $SU(2)_L$ gauge-singlet fields: a charged scalar ($S$) and a right-handed fermion ($N_R$). The quantum numbers of these new states are shown below

\begin{table}[!h]
\setlength\tabcolsep{12pt}
\begin{center}
    \begin{tabular}{c c c c c}
    \toprule 
     Field    & $SU(3)_c$ & $SU(2)_L$ & $U(1)_Y$ & $Z_2$  \\ 
     \toprule
     $S$   & {\bf 1}   &  {\bf 1}  & $+2$   &  $-1$   \\
     $N_R$ & {\bf 1}   &  {\bf 1}  & $0$    &  $-1$   \\
     \toprule
    \end{tabular}
    \caption{The new particles of the model and their representation under $SU(3)_c \otimes SU(2)_L \otimes U(1)_Y \otimes Z_2$.}
\end{center}
\end{table}

In this setup, we have introduced an accidental discrete symmetry ($Z_2$) under which the new particles are odd while all the SM particles are even. In this case, the right-handed fermion, being neutral particle, is a suitable candidate for DM, assuming it is lighter than the singlet scalar.  We furthermore assume that the charged scalar singlet carries a lepton number which therefore implies that it only interacts with the SM charged leptons (right-handed leptons in particular). The resulting interaction would be similar to the case of slepton-neutralino-lepton in supersymmetric theories with the exception that here we only have one {\it slepton} particle that interacts with all the lepton generations. Under these assumptions, the most general Lagrangian is given by 
\begin{eqnarray}
{\cal L} = {\cal L}_S + {\cal L}_N + {\cal L}_{SN} - V(\Phi, S),
\end{eqnarray}
where ${\cal L}_S$ represents the kinetic term for the charged scalar singlet, given by
\begin{eqnarray}
    {\cal L}_S &=& (\mathcal{D}^\mu S)^\dagger(\mathcal{D}_\mu S) = (\partial^\mu S)^\dagger (\partial_\mu S) - (e A^\mu - e \tan\theta_W Z^\mu) S^\dagger \overset{\leftrightarrow}{\partial}_\mu S \nonumber \\ 
 &+& e^2 A_\mu A^\mu S^\dagger S
 + e^2 \tan^2\theta_W Z_\mu Z^\mu S^\dagger S - 2 e^2 \tan\theta_W A_\mu Z^\mu S^\dagger S,
 \label{eq:S:gauge}
\end{eqnarray}
Here, in the first line, the first term refers to the kinetic energy of $S$, while the second term represents the triple interaction of $S$ with the photon and the $Z$ boson. The second line illustrates the quartic interaction of $S$ with the $\gamma/Z$. In equation \ref{eq:S:gauge}, we have $e = \sqrt{4 \pi \alpha_{\rm EM}}$ as the electric charge, $\theta_W$  denoting the Weinberg weak mixing angle, and $A \overset{\leftrightarrow}{\partial_\mu}B \equiv A(\partial_\mu B) - (\partial_\mu A)B$. The notation ${\cal L}_N + {\cal L}_{SN}$ refers to the Lagrangian of the right-handed fermion and its interaction with $S$, which can be expressed as follows:
\begin{eqnarray}
    {\cal L}_N + {\cal L}_{SN} \equiv i \bar{N}_R \slashed{\partial} N_R^c + \frac{1}{2} M_N \bar{N}_R N_R^c + \bigg(Y_\ell \bar{\ell}^c_R N_R S + {\rm h.c.} \bigg),
\end{eqnarray}
where $M_N$ is the mass of the $N_R$ particle and $Y_{\ell, \ell=e,\mu,\tau}$ are assumed to be real-valued couplings. In the last term, the sum over the lepton generations is implicit. Finally, the scalar potential is given by 
\begin{equation}
V(\Phi, S) = - m_{11}^2 |\Phi^\dagger \Phi| + m_{22}^2 |S^\dagger S| + \lambda_1 |\Phi^\dagger \Phi|^2 + \lambda_2 |S^\dagger S|^2 + \lambda_3 |\Phi^\dagger \Phi| |S^\dagger S|,
\end{equation}
where $\Phi$ is the SM Higgs doublet:
$$
\Phi \equiv 
\left(\begin{array}{c}
G^+ \\
\frac{1}{\sqrt{2}} (\upsilon + h + i G^0)
\end{array}
\right),
$$
and $\upsilon = \sqrt{2 G_F}$ representing the vacuum expectation value (VEV), while $G^0$ and $G^+$ correspond to the Nambu-Goldstone bosons responsible for forming the longitudinal polarizations of the $Z$ and $W$ bosons, respectively. Following the electroweak symmetry breaking, we are left with three scalars: the CP-even scalar, identified as the 125 GeV Higgs boson, and a pair of charged scalars denoted as $H^\pm$ in the subsequent discussion. Their tree-level masses are determined by: 
\begin{eqnarray}
M_{H_{\rm SM}}^2 = \lambda_1 \upsilon^2 = - 2 m_{11}^2, \quad M_{H^\pm}^2 = m_{22}^2 + \frac{1}{2} \lambda_3 \upsilon^2.
\end{eqnarray}
In addition to the SM parameters, this model has seven more parameters defined by
\begin{eqnarray}
\{M_{H^\pm}, M_{N_R}, \lambda_2, \lambda_3, Y_e, Y_{\mu}, Y_{\tau} \}.
\label{eq:params}
\end{eqnarray}
The model is subject to various theoretical and experimental constraints. In this study, we take into account constraints from the stability of the scalar potential, unitarity of the scattering amplitudes, bounds from Higgs boson decays and lepton flavour violation. Additionally,  we account for the constraints associated with the dark matter relic density and the direct detection experiments. Finally recasting the searches of sleptons and neutralinos at the LHC implies that mediator masses less than $400$ GeV are excluded. For more details, we refer the interested reader to Refs. \cite{Jueid:2020yfj,  Jueid:2023abcs}. 

\subsection{Benchmark points and scanning procedure}

\begin{table*}[!t]
\setlength\tabcolsep{8pt}
\begin{center}
\begin{adjustbox}{max width=0.92\textwidth}
\begin{tabular}{lcccc}
\toprule
\multicolumn{1}{c} { Benchmark scenario } & BP1 & BP2 & BP3 & BP4 \\
\midrule
\multicolumn{5}{l}{\textit{Parameters}} \\
\midrule
$M_{N_R}~({\rm GeV})$    & $50$  &   $200$ & $598$ & $1000$ \\
$M_{H^\pm}~({\rm GeV})$ & $500$ & $500$ & $600$  & $1500$  \\
$Y_{e}$       & $10^{-4}$ &     $5 \times 10^{-4}$ & $10^{-3}$ & $5 \times 10^{-3}$ \\
$Y_{\mu}$   & $2.8$      &     $1.6$  & $1$ & $2$  \\
$Y_{\tau}$   & $5 \times 10^{-2}$  & $5 \times 10^{-1}$ & $5 \times 10^{-1}$ & $2$  \\
$\lambda_3$  & $4$         &      $5$ & $5$ & $6$  \\
\bottomrule
\end{tabular}
\hspace{0.2cm}
\end{adjustbox}
\end{center}
\caption{Benchmark points used for the differential distributions in this analysis. More details can be found in Ref. \cite{Jueid:2023abcs}.}
\label{tab:BSs}
\end{table*} 

We will start by examining  four benchmark points, as shown in table \ref{tab:BSs} and have been  previously considered in Ref. \cite{Jueid:2023abcs}. The illustration of the differential distributions and the performance of our algorithms will be discussed in great detail for these benchmark scenarios. Interestingly, the model predicts very simple relations between physical observables -- production cross sections, relic density and spin-independent cross sections -- and model parameters as follows:
\begin{eqnarray}
    \sigma_{\mu \mu \to N_R N_R H_{\rm SM}} &\propto& \lambda_3^2 \times Y_\mu^4 \times {\cal F} (s, M_{H^\pm}^2, M_{N_R}^2), \nonumber \\
    \Omega_{N_R} h^2 &\propto& Y_\mu^{-4} \times {\cal G} (M_{H^\pm}^2, M_{N_R}^2), \\ 
    \sigma_{\rm SI} &\propto& \lambda_3^2  \times Y_\mu^4 \times {\cal H} (M_{H^\pm}^2, M_{N_R}^2), \nonumber 
    \label{eq:scaling}
\end{eqnarray}
where ${\cal F}$, ${\cal G}$ and ${\cal H}$ are functions of $M_{N_R}^2$ and $M_{H^\pm}^2$. Therefore, it is clear that the physical observables exhibit simple scaling properties when changing from one choice of the parameters $Y_\mu$ and $\lambda_3$ to another choice. We therefore perform the following scan over the mass of the DM particle of the model
\begin{eqnarray}
    50 \leq M_{N_R} \leq \frac{1}{2} \bigg(\sqrt{s} - M_{H_{\rm SM}} \bigg),
    \label{eq:scan}
\end{eqnarray}
for $\Delta = 200$ GeV where $\Delta = M_{H^\pm} - M_{N_R}$. We furthermore fix $Y_\mu = 2 = \lambda_3$\footnote{Note that the choice of the other couplings $Y_e$ and $Y_\tau$ does not affect the results of our analysis.}. 

\section{Technical setup}
\label{sec:technics}

\subsection{Signal and backgrounds}
\label{sec:SandB}

\begin{figure*}[!t]
\centering
\includegraphics[width=0.6\linewidth]{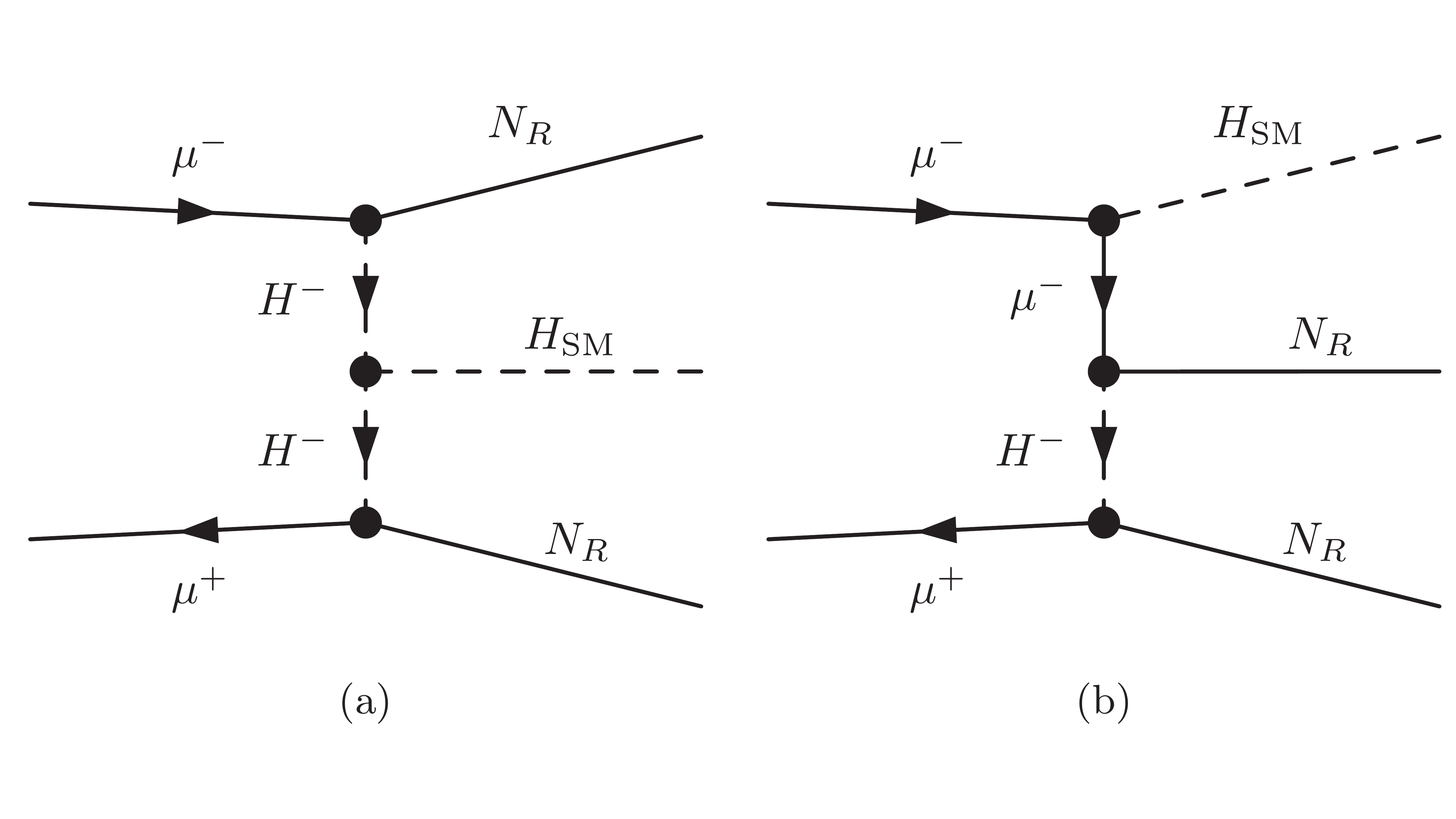}
\caption{Example of Feynman diagram for the signal process $\mu\mu \to N_R N_R H_{\rm SM}$. }
\label{fig:FD:signal}
\vfill
\includegraphics[width=0.8\linewidth]{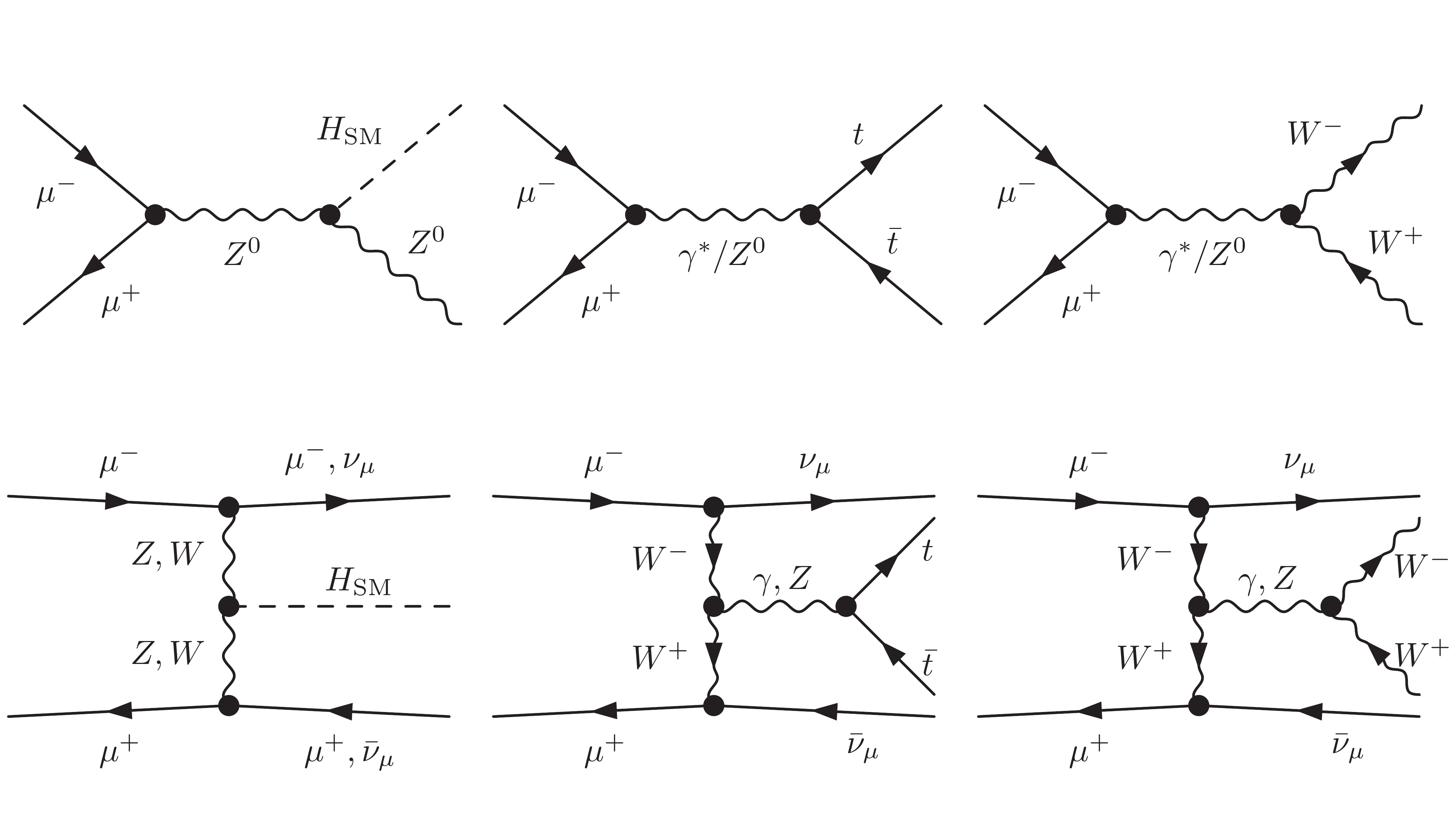}
\vspace{-0.5cm}
\caption{Example of Feynman diagrams for the background processes contributing to the $b\bar{b}+E_{T}^{\rm miss}$ final state. Here we show the muon-annihilation channels (upper panel) and VBF channels (lower panel).}
\label{fig:FD:backgrounds}
\end{figure*}

\begin{table}[!t]
\setlength\tabcolsep{8pt}
\begin{center}
\begin{adjustbox}{max width=\textwidth}
\begin{tabular}{l l c c c}
\toprule
& & & & \\ [-0.4ex]
 &  $\sqrt{s_{\mu\mu}}$ [TeV] &  $3$ & $10$ & $30$ \\ [2.3ex]
\toprule
& & & &  \\ [-0.2ex]
\multirow{4}{*}{$\sigma \times {\rm BR}~[{\rm fb}]~(N_{\rm events})$} & BP1 & $6.70 \times 10^{1}~(6.70\times 10^4)$ & $1.02 \times 10^0~(1.02\times 10^4)$ & $3.67 \times 10^{-2}~(3.67 \times 10^3)$  \\ 
& BP2 & $1.02 \times 10^1~(1.02\times 10^4)$ & $0.31 \times 10^0~(0.31 \times 10^4)$  & $1.12 \times 10^{-2}~(1.12\times 10^3)$  \\
& BP3 & $0.82 \times 10^0~(0.82 \times 10^3)$ & $0.47 \times 10^{-1}~(0.47\times 10^3)$  & $1.81 \times 10^{-3}~(1.81 \times 10^2)$  \\ 
& BP4 & $0.11 \times 10^0~(0.11 \times 10^3)$ & $0.21 \times 10^0~(0.21 \times 10^4)$  & $1.47 \times 10^{-2}~(1.47 \times 10^3)$ \\ [0.8ex]
\bottomrule
\end{tabular}
\hspace{0.2cm}
\end{adjustbox}
\end{center}
\caption{\label{tab:FS:NNH} 
The total cross sections times the branching ratio ($\sigma \times {\rm BR}$) and the expected number of signal events for the $N_R N_R H_{\rm SM}~(\to b\bar{b})$. We consider three representative center-of-mass energies of $3$, $10$ and $30$ TeV and we show the results for the benchmark points defined in table \ref{tab:BSs}.}
\end{table}

In this work, we investigate the potential discovery of dark matter at muon colliders using  the mono-Higgs channel: $N_R N_R H_{\rm SM}(\to b\bar{b})$. Our  analysis focuses  on collision energy $\sqrt{s} = 3$ TeV and an integrated luminosity of $1~{\rm ab}^{-1}$. The $N_R N_R H_{\rm SM}(\to b\bar{b})$ process leads to a final state comprising of missing energy and at least two $b$-tagged jets (in the resolved case) or at least one large-$R$ jet (in the boosted regime). The parton-level Feynman diagrams for the signal and the backgrounds are depicted in figures \ref{fig:FD:signal} and \ref{fig:FD:backgrounds}. The signal cross section receives two contributions, both occurring in $t$-- and $u$--channels. The first contribution occurs through the double exchange of charged singlet scalar (\ref{fig:FD:signal}-a) while the second contribution occurs through the exchange of a muon in the $t$--channel (\ref{fig:FD:signal}-b). The second contribution is negligible since it is suppressed by the smallness of the Higgs-muon Yukawa coupling. The backgrounds can be split in two categories depending on their exact signature at the parton level: 

\begin{itemize}
    \item Irreducible backgrounds: This category of background involves either the production of the SM Higgs boson in association with two SM neutrinos or the production of two gauge bosons ($ZZ/WZ/WW$) where one gauge boson decays hadronically while the other decays invisibly (in the case of the $Z$--boson) or leptonically (in the case of $W$--boson) with one charged lepton escapes the detection. Note that the diboson production can be significantly reduced by requirements on the invariant of the hadronically decaying gauge boson to be off their on-shell mass window.   

    \item Reducible backgrounds: This category contains the production of $t\bar{t}$ and $t\bar{t}+W/Z$ or the production through neutral current VBF (i.e. involving two charged leptons). We note that this category can be significantly reduced by several requirements on the number of hard charged leptons, or requirements on the invariant mass of the $b\bar{b}$ system that will form a Higgs {\it candidate}.  
\end{itemize}

\begin{figure*}[!t]
    \centering
    \includegraphics[width=0.495\linewidth]{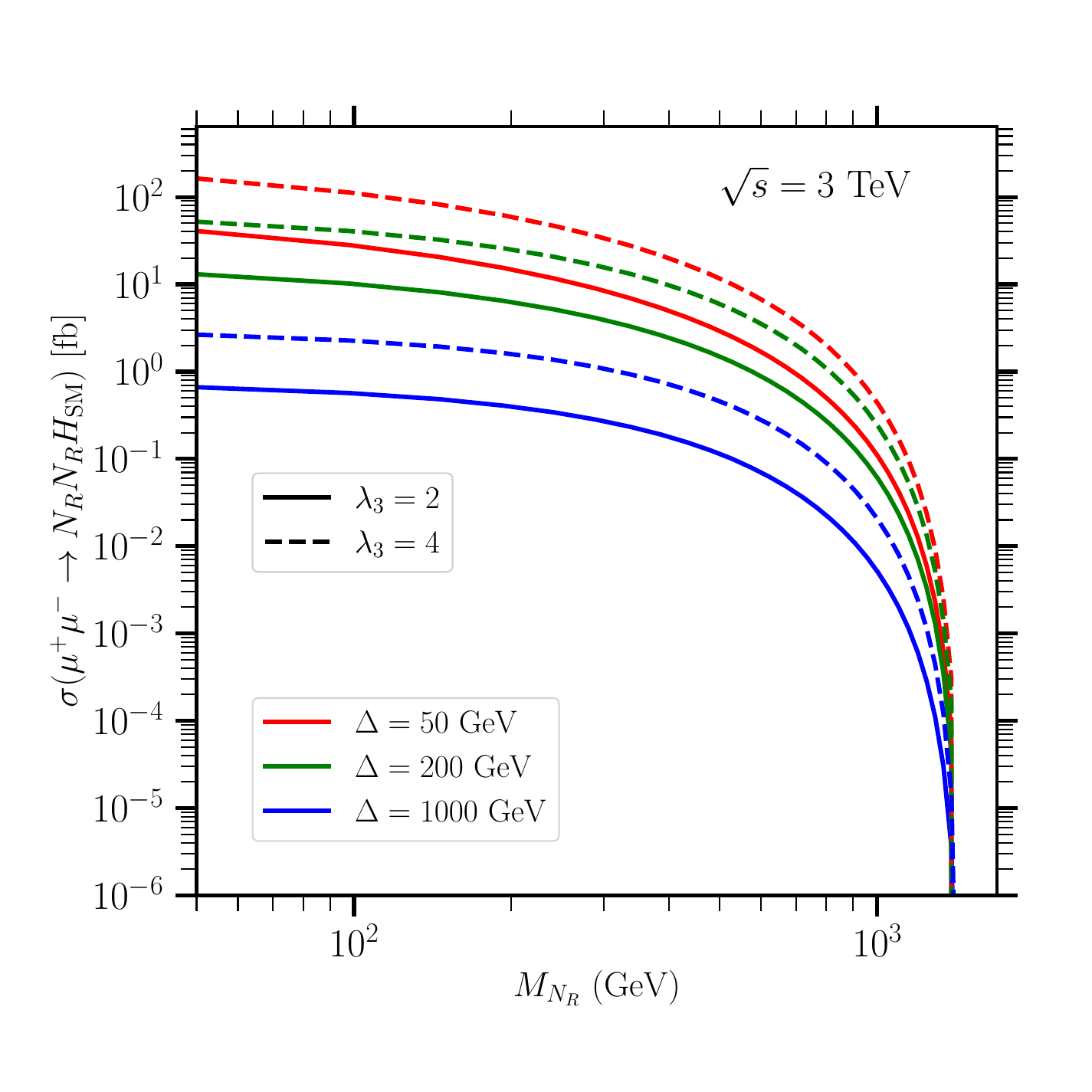}
    \hfill 
    \includegraphics[width=0.495\linewidth]{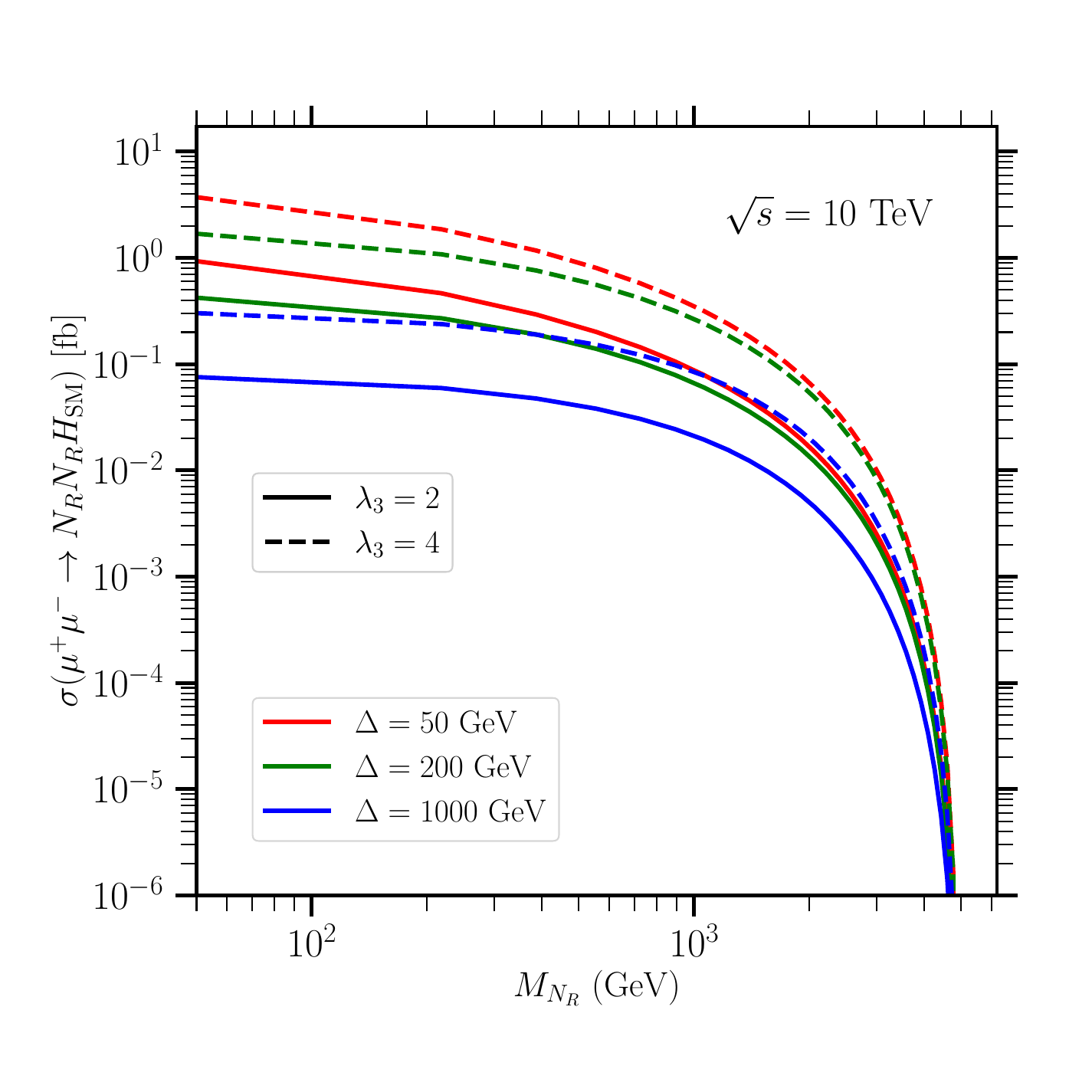}
    \vspace{-0.7cm}
    \vfill 
    \includegraphics[width=0.495\linewidth]{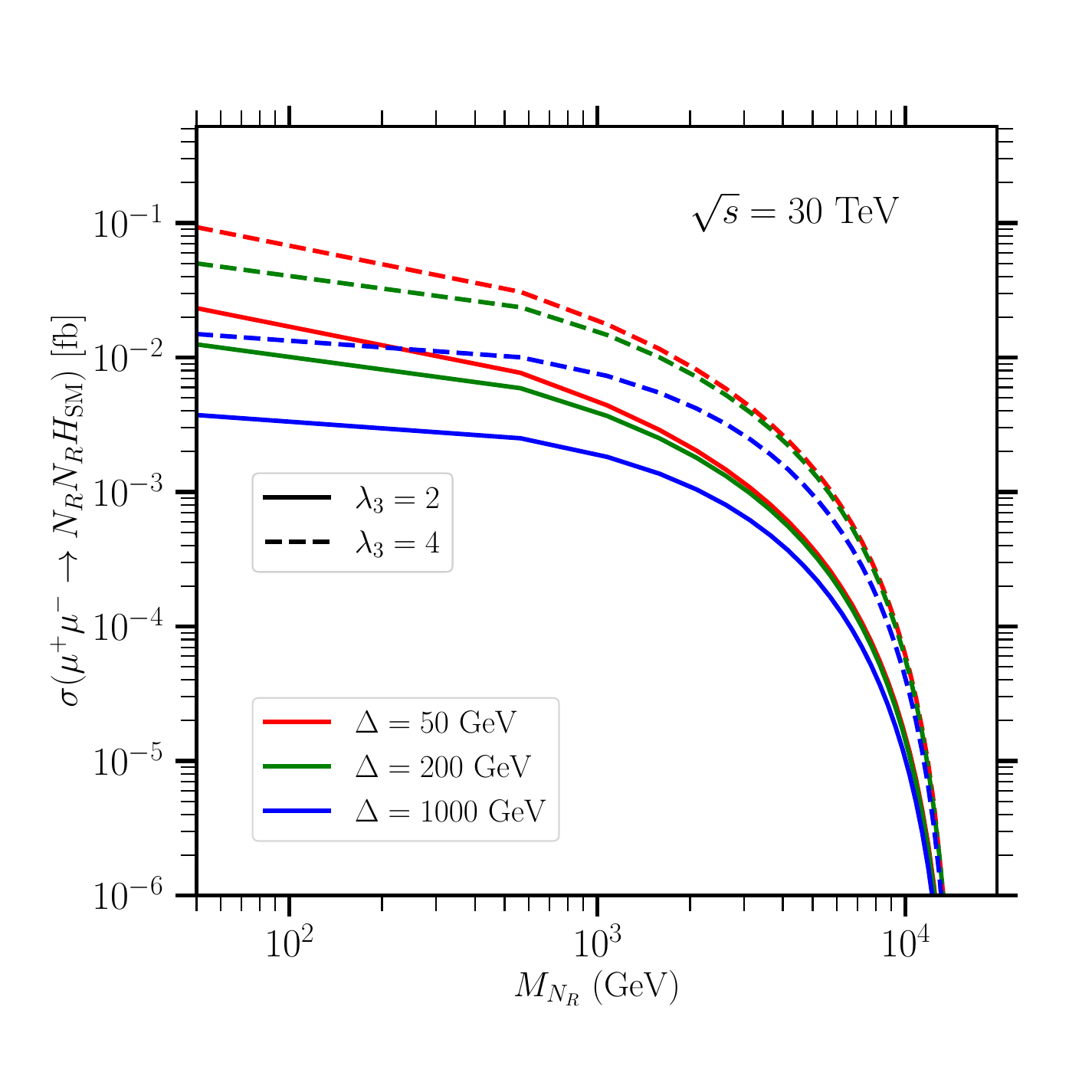}
    \vspace{-0.7cm}
    \caption{The cross section for the production of $N_R N_R H_{\rm SM}$ as a function of $M_{N_R}$ for $\sqrt{s}=3$ TeV (left upper panel), $\sqrt{s} = 10$ TeV (right upper panel) and $\sqrt{s} = 30$ TeV (lower panel). Here we show the results for $\lambda_3 = 2$ (solid lines) and $\lambda_3 = 4$ (dashed lines) for $\Delta = 50$ GeV (red), $\Delta = 200$ GeV (green) and $\Delta = 1000$ GeV (blue). All the calculations were done for $Y_\mu = 2$.}
    \label{fig:XS:NNH}
\end{figure*}

We present the cross-section results for the signal process in Table \ref{tab:FS:NNH} corresponding to the four benchmark points defined in Table \ref{tab:BSs} and shown in Figure \ref{fig:XS:NNH}, as a function of the DM mass ($M_{N_R}$). The background cross sections are shown in table \ref{tab:XS:backgrounds}.

\begin{table}[!t]
\setlength\tabcolsep{24pt}
\begin{center}
\begin{adjustbox}{max width=1.01\textwidth}
\begin{tabular}{lccc}
\toprule
\multicolumn{1}{c} { Center-of-mass energy } & $3$ TeV & $10$ TeV & $30$ TeV \\
\midrule
\multicolumn{4}{c}{\textit{Cross section~(fb)}} \\
\midrule
$\mu\mu \to H_{\rm SM}Z$ & $1.37 \times 10^0$ & $0.12 \times 10^0$ & $1.37 \times 10^{-2}$  \\
$\mu\mu \to WW$ & $4.67 \times 10^2$ &	$5.89 \times 10^1$ &  $8.26 \times 10^0$ \\
$\mu\mu \to ZZ$ & $2.61 \times 10^1$ & 	$3.28 \times 10^0$ &  $4.60 \times 10^{-1}$ \\
$\mu\mu \to t\bar{t}$ & $1.91 \times 10^1$ &	$1.72 \times 10^0$  &	$1.92 \times 10^{-1}$ \\
$VV \to HZ$ & $9.87 \times 10^0$ 	& $3.53 \times 10^1$  &	$7.59 \times 10^1$  \\
$VV \to H_{\rm SM}$ & $4.98 \times 10^2$ & $8.45 \times 10^2$ & 	$1.17 \times 10^3$ \\
$VV \to WZ$ & $3.98 \times 10^1$ & $3.19 \times 10^1$ &  $1.26 \times 10^1$\\
$VV \to WW$ & $1.51 \times 10^2$ & $4.30 \times 10^2$ &   $8.58 \times 10^2$\\
$VV \to ZZ$ & $5.66 \times 10^1$ &  $2.03 \times 10^2$ &  $4.30 \times 10^2$ \\
$VV \to t\bar{t}$ & $5.22 \times 10^0$ 	& $1.71 \times 10^1$ &	$3.14 \times 10^1$ \\
$VV \to t\bar{t}W$ & $5.67 \times 10^{-2}$ &  $1.05 \times 10^{-1}$ & 	$6.97 \times 10^{-2}$  \\
$VV \to t\bar{t}Z$ & $1.10 \times 10^{-1}$ & $9.01 \times 10^{-1}$	& $2.77 \times 10^0$ \\
\bottomrule
\end{tabular}
\hspace{0.2cm}
\end{adjustbox}
\end{center}
\caption{Parton-level cross sections for the background processes that we are taken into account in this study.}
\label{tab:XS:backgrounds}
\end{table} 

\subsection{Monte Carlo event generation}
\label{sec:MC}

Samples for the signal and the backgrounds were generated using \textsc{MadGraph\_aMC@NLO} version 3.4.1 \cite{Alwall:2014hca} where we have used a dedicated model file in the UFO format \cite{Degrande:2011ua} which we have produced using \textsc{FeynRules} version 2.3.0 \cite{Alloul:2013bka}. The model file along with instructions on how to use it can be found in this \href{https://feynrules.irmp.ucl.ac.be/wiki/MinimalLeptonPortalDM}{link}. For the generation of both the  signal and the background events, we have imposed some generator-level cuts:

\begin{eqnarray*}
    p_T^\ell > 7~{\rm GeV}~&{\rm and}&~|\eta_\ell| < 7 \\
    p_T^{j} > 10~{\rm GeV}~&{\rm and}&~|\eta_j| < 6.
\end{eqnarray*}

Background processes, on the other hand, receive sizeable contributions from the VBF channels, {\it i.e.} $VV \to X$. The computation of the rates of these processes can be done by either considering the VBF/VBS of two gauge bosons through $\mu^+ \mu^-$ process following the lines of Ref. \cite{Costantini:2020stv} or considering that electroweak gauge bosons form partons within muons with some parton distribution functions (PDFs) within the muons \cite{Ruiz:2021tdt}. Given that there is no validated treatment of initial state gauge boson PDFs for parton showers, we use the first approach in our simualtion of the VBF processes in the SM. For example, to simulate $\mu^+ \mu^- \to W^* W^* \to t\bar{t} + X$, we use the following syntax in \textsc{MadGraph5\_aMC@NLO}:
\begin{verbatim}
    > import model sm
    > generate mu+ e- > t t~ vm~ ve
    > add process mu+ e- > t t~ mu+ e-
    > output MyOutput
\end{verbatim}
This syntax is necessary to isolate VBF contributions from the corrections of initial-state radiation (ISR) or final-state radiation (FSR) to $s$--channel contributions. Note that the syntax above corresponds to $\mu^+ e^-$ scatterings and similar results can be found if we consider the complex conjugate of it, {\it i.e.} $\mu^- e^+$ scatterings. In our calculations, we have considered both neutral-current as well as charged-current contributions to the VBF processes. For example, $VV \to t \bar{t}$ in table \ref{tab:XS:backgrounds} includes both contributions from $\mu^+ e^- \to t\bar{t} \bar{\nu}_\mu \nu_e$ and $\mu^+ e^- \to t\bar{t} \mu^+ e^-$. We have checked that our calculations of the background cross sections yield excellent agreement with the results of ref. \cite{Costantini:2020stv}. As for the signal processes, we found that there is no contribution to the production of the $N_R$ through VBF. This is can be understood since the $N_R$ particles are $SU(2)_L \otimes U(1)_Y$ singlets and therefore do not couple directly to $\gamma/Z$ or $W$ gauge bosons. For each benchmark point in the signal and for all the backgrounds we have generated about $9 \times 10^{5}$--$3 \times 10^6$ parton-level events. The produced events are passed to \textsc{Pythia} version 8307 \cite{Bierlich:2022pfr} to add resonance decays, parton showering and hadronisation.

\subsection{Object definitions}
\label{sec:objects}

In this section, we discuss the object definitions at the reconstruction level that we have used in our analysis. In this work, we define charged leptons (electrons or muons), hadronically decaying tau leptons ($\tau_h$), small--$R$ jets, large--$R$ jets and missing transverse momentum. The details are shown below:

\begin{itemize}

\item \textbf{Electron candidates}: Are required to have $p_T^e > 7$ GeV and $|\eta| < 6$. 

\item \textbf{Muon candidates}: Muon candidates are required to have $p_T > 7$ GeV and $|\eta| < 6$. 

\item \textbf{Hadronically-decaying tau leptons}: Those are reconstructed from their one-, two- and three-prong decays and are required to have $p_T > 15$ GeV and $|\eta_\tau| < 2.5$. We assume a $\tau$--tagging efficiency of $60\%$~($70\%$) if the number of associated tracks is $\geq 2$~($1$). 

\item \textbf{Small-$R$ Jets}: Candidate jets are reconstructed with the anti--$k_t$ algorithm with a radius parameter $R = 0.4$ which are referred to as ``small-R jets'' \cite{Cacciari:2008gp}. These jets are further required to have $p_T > 25$ GeV and $|\eta_j| < 6$. We use a ghost-based approach to tag ``small--$R$ jets'' as $b$-jets where we assume a $70\%$ $b$-tagging efficiency and a $p_T$-dependent mistagging efficiency of light and charm jets as $b$-jets, {\it i.e.}
\begin{eqnarray}
    {\cal E}_{c|b} &=& 0.20~{\rm tanh}(0.02 p_T) \frac{1}{1 + 0.0035 p_T} \nonumber \\
    {\cal E}_{j|b} &=& 0.002 + 7.3 \times 10^{-6} p_T.
\end{eqnarray}
With these parameterisations one gets ${\cal E}_{j|b} = 0.002~(0.009)$ and ${\cal E}_{c|b} = 0.085~(0.045)$ for $p_T = 25~(1000)$ GeV. $b$-tagged jets are required to have $p_T > 25$ GeV and $|\eta_b| < 2.5$.

\item \textbf{Large--$R$ Jets}: In our analysis, the SM Higgs boson can be produced with very high transverse momentum. In such cases, the hadronic decay products of the SM Higgs boson can not be resolved into two isolated jets and therefore clustering based on small--$R$ jet radius will have very small efficiencies. Therefore, we also utilize large--$R$ jets in our analysis. A large radius parameter is chosen in order for a single large--$R$ jet to capture all the constituents that are produced in the decay of a boosted Higgs boson. We perform two independent clustering algorithms along the same lines of the ATLAS \cite{ATLAS:2021shl} and CMS \cite{CMS:2018zjv} analyses. First, we cluster jets using the anti--$k_t$ clustering algorithm with a jet radius of $R=1$ \cite{Cacciari:2008gp} and these jets will be labeled as AK10 jets. Furthermore, we apply a trimming algorithm to remove any soft radiation \cite{Krohn:2009th}. For this purpose we use the $k_t$ algorithm \cite{Catani:1993hr} where we remove any subjet of radius $R=0.2$ that carries less than $5\%$ of the total AK10 jet energy. An independent clustering algorithm will be used in our analysis where we cluster the jets using the Cambridge-Aachen algorithm and a jet radius of $R=1.5$ \cite{CMS:2009lxa} which will be denoted as CA15 jets. In this case, the soft-drop jet grooming algorithm \cite{Larkoski:2014wba} is employed to cut soft and wide-angle radiation from the CA15 jets where we use $\beta = 1$ and $z_{\rm cut} = 0.1$. For the training of the signal and the backgrounds and in order to optimise the sensitivity of the analysis, we use the ratios of the energy correlation functions constructed from the output of the CA15 jets. In this analysis, we construct two variables $N_2$ and $M_2$ which were found to be very powerful in discriminating two-prong boosted objects from QCD jets or three-prong boosted jets \cite{Moult:2016cvt}. They are defined as 
\begin{eqnarray}
    M_2^{(\beta)} = \frac{_1e_{3}^{(\beta)}}{_1e_2^{(\beta)}}, \quad
    N_2^{(\beta)} = \frac{_2e_{3}^{(\beta)}}{(_1e_2^{(\beta)})^2},
\end{eqnarray}
where $_k e_{i}^{(\beta)}$ are the generalized $i$-point energy correlation function for the $k$ pair-wise angles that enters their products and $\beta$ is a parameter that controls the overall angular scaling of these operators. In this analysis, we choose $\beta$ to be equal to $1$. All the large--$R$ jets are required to have $p_T > 200$ GeV and $|\eta| < 2.5$. 

\item \textbf{Missing Transverse Energy}: The missing transverse momentum $\bf{p}_T^{\rm miss}$ (with magnitude $E_T^{\rm miss}$) is the  negative vector sum of the $p_T$ of all selected and calibrated objects in the event, including a term to account for energy from soft particles in the event which are not associated
with any of the selected objects.

\end{itemize}

We further impose isolation requirements on charged leptons. A lepton isolation criterion is defined by imposing a cut on the following quantiy
$$
I_R \equiv \sum_{i~\in~{\rm tracks}} p_T^{i},
$$
where the sum includes all tracks (excluding the lepton candidate itself) within the cone defined by $\Delta R < R_{\rm cut}$ about the direction of the charged lepton. The value of $R_{\rm cut}$ is the smaller of $r_{\rm min}$ and $10~{\rm GeV}/p_T^\ell$, where $r_{\rm min}$ is set to $0.3$ for both the electron and the muon
candidates, and $p_T^\ell$ is the lepton transverse momentum. All the charged lepton candidates must satisfy 
$$
I_R /p_T^\ell < 0.3,
$$
which defines a loose-isolation criterion. Overlap removals are used to remove leptons or jets if they are within some defined $\Delta R$ of a given object. Electron~(muon) candidates that lie within $\Delta R = 0.2~(0.4)$ of a jet candidate. Jets are also required to have $\Delta R = 0.4$ of other leptons or jets in the event. \\

Fast detector simulation is performed using the \texttt{SFS} module \cite{Araz:2020lnp} in \textsc{MadAnalysis}~5 \cite{Conte:2012fm,Dumont:2014tja,Conte:2014zja,Conte:2018vmg,Araz:2019otb,Araz:2021akd}. The calculation of the jet observables including the clustering of large--$R$ jets and the removal of the soft radiation is done with the help of customised \texttt{C++} analysis within the \texttt{Substructure} module \cite{Araz:2023axv}. All the jets were clustered using \textsc{FastJet} version 3.4.0 \cite{Cacciari:2011ma}. The momentum smearing and identification efficiencies for electrons are implemented from the detector design of FCC--hh that is shipped in \textsc{Delphes}~3.4.0 \cite{deFavereau:2013fsa} (can be found in \url{https://github.com/delphes/delphes/blob/master/cards/delphes_card_MuonColliderDet.tcl}). For muons smearing and identification efficiencies we use the results of Ref. \cite{MuonCollider:2022ded}.

\section{Cut-based analysis}
\label{sec:cut}

In this section we discuss the basic approach for the signal-to-background analysis which consists of performing basic event selection on both the signal and the background processes. In  this analysis, we follow the analysis strategies of the recent \textsc{ATLAS} \cite{ATLAS:2021shl} and \textsc{CMS} \cite{CMS:2018zjv} searches of DM produced in association with a Higgs boson and decaying to bottom quarks. 

\subsection{Resolved regime}

\begin{figure}[!t]
    \centering
    \includegraphics[width=0.329\linewidth]{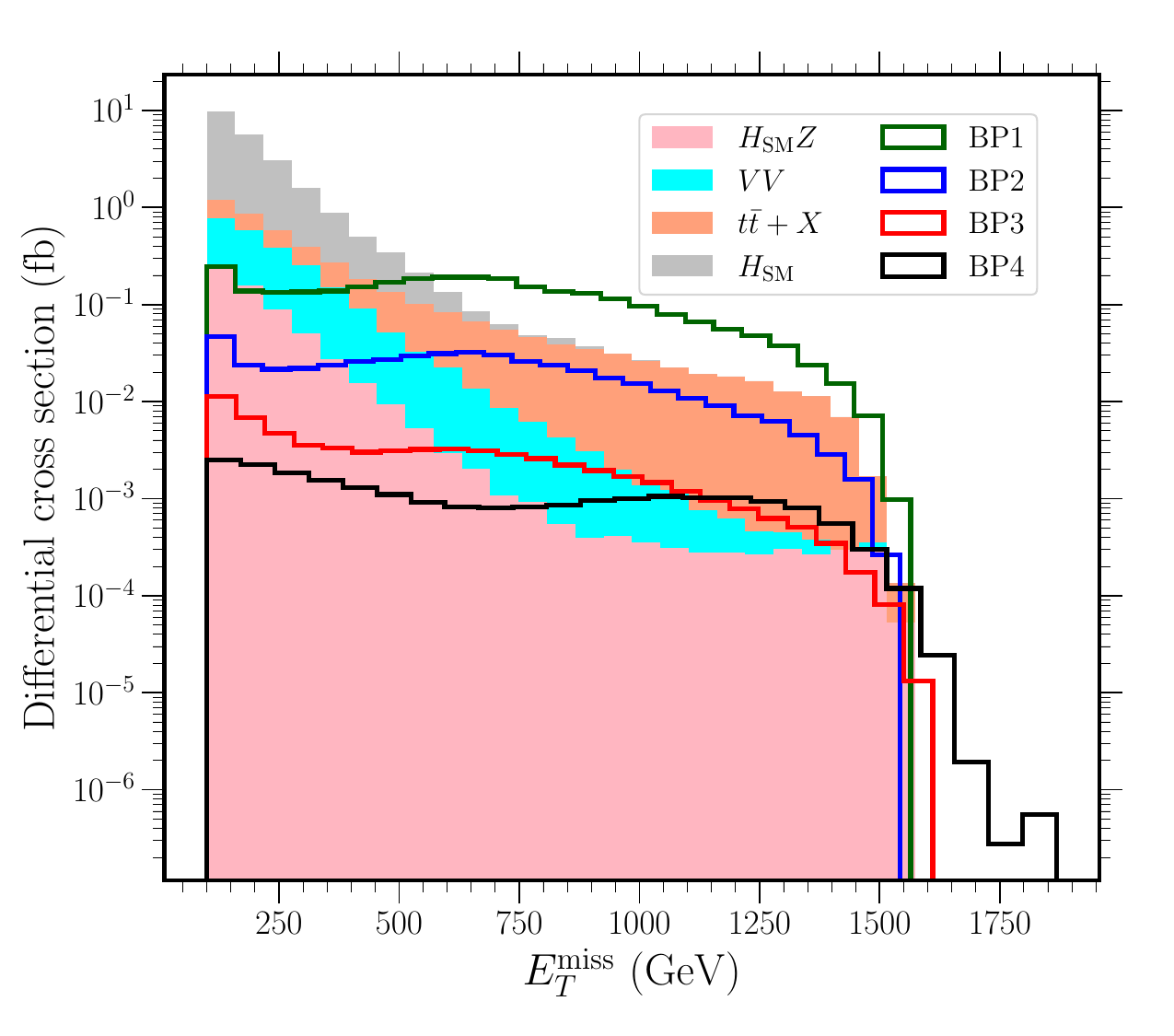}
    \hfill  
    \includegraphics[width=0.329\linewidth]{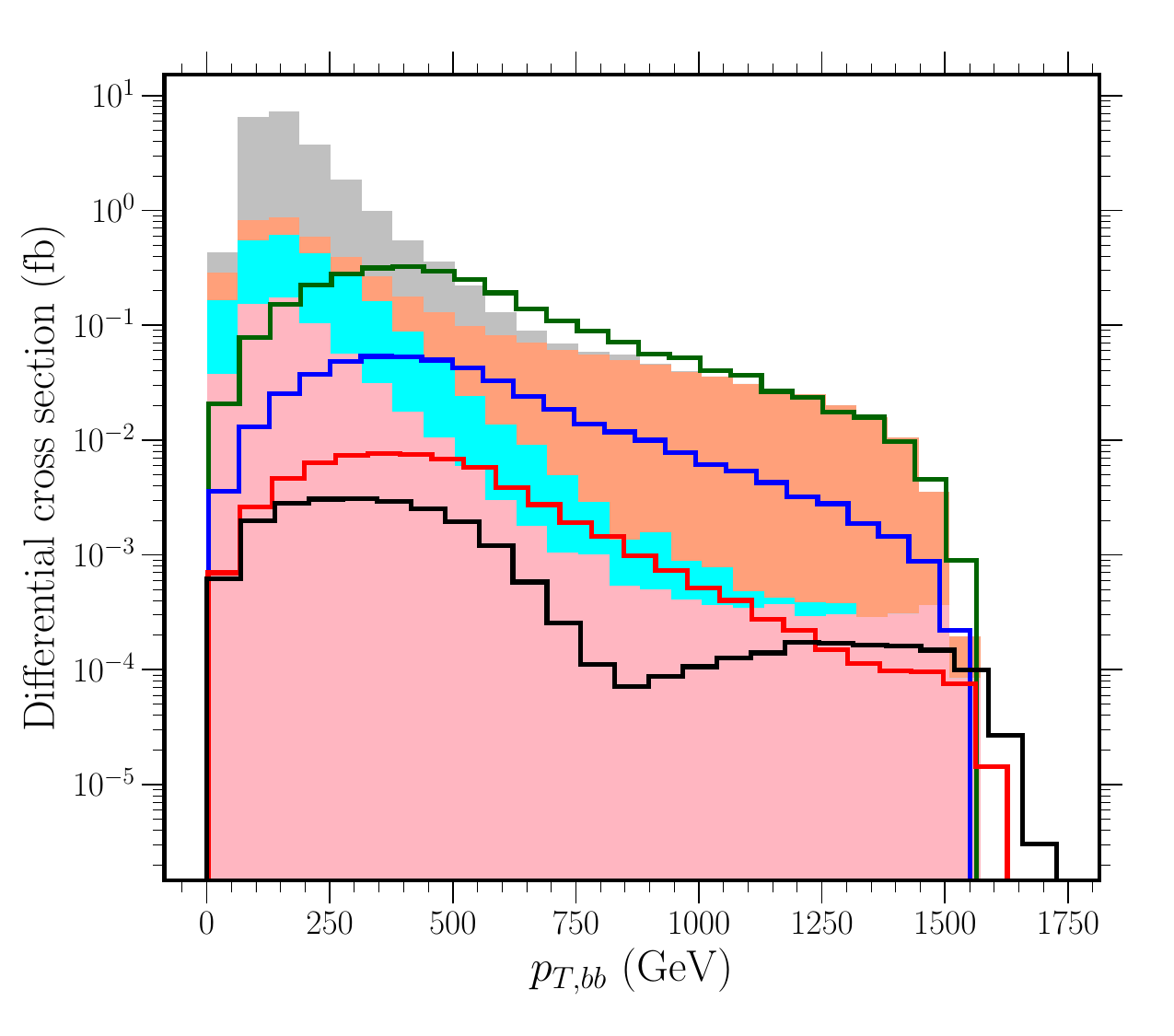}
    \hfill
    \includegraphics[width=0.329\linewidth]{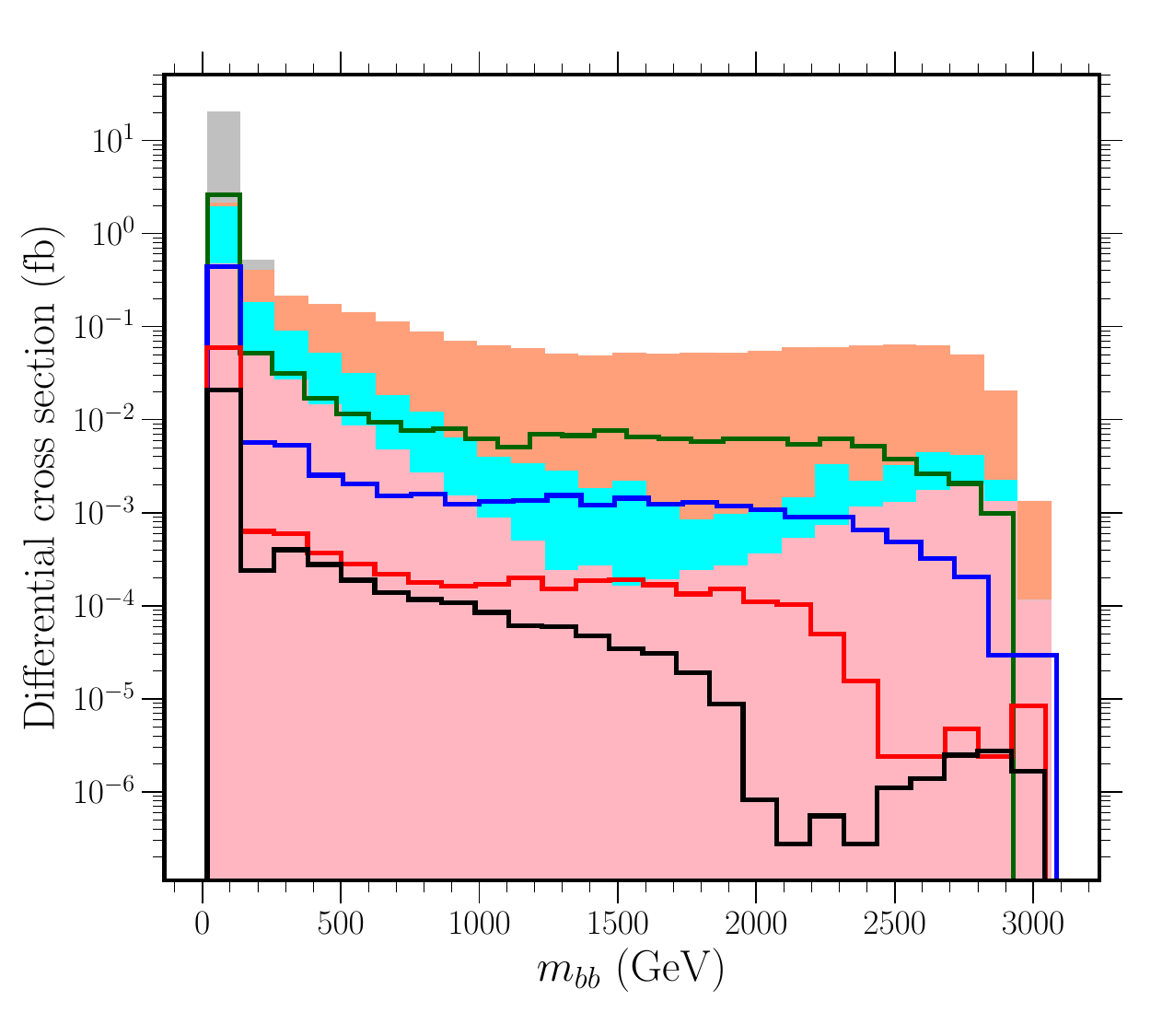}
    \caption{Differential cross section per bin for the four benchmark points defined in Table \ref{tab:BSs} and the background processes shown as stacked histograms in the resolved regime. From the left to the right, we show the missing transverse energy, the transverse momentum of the Higgs candidate and, the invariant mass of the Higgs candidate. More details can be found in the text.}
    \label{fig:distributions:resolved}
\end{figure}

For the resolved regime, we assume that the Higgs boson candidate is reconstructed from two small-$R$ well-separated $b$-tagged jets. The key distributions for the signal-to-background optimisations are shown in Fig. \ref{fig:distributions:resolved}. First, we require that the missing transverse energy is larger than $100$ GeV. Electrons and muons that pass the loose isolation criteria defined in the previous section are vetoed. This criterion reduces the VBF processes that occur through neutral current interactions, {\it i.e.} $Z^*Z^* \to X$. Events that contain at least one hadronically-decaying $\tau$--lepton satisfying $p_T^\tau > 15$ GeV and $|\eta^\tau| < 6$ are vetoed. We do not require the existence of any photon with $p_T > 10$ GeV and $|\eta| < 6$. We furthermore require that the events contain at least three small-$R$ untagged jets that satisfy $p_T > 25$ GeV and $|\eta| < 6$ where at least two of them are $b$--tagged with $|\eta^b| < 2.5$. To reduce backgrounds where $E_{T}^{\rm miss}$ arose from mismeasurement of the jet $p_T$  or from leptonic decays of heavy flavours, we require that the minimum $\Delta\phi$ defined by 
$$
\Delta\phi_{\rm min} \equiv {\rm min}\bigg\{\Delta\phi({\bf p}_{\rm miss}, {\bf j_1}), \Delta\phi({\bf p}_{\rm miss}, {\bf j_2}), \Delta\phi({\bf p}_{\rm miss}, {\bf j_3})\bigg\},
$$
to be larger than $20^{\circ}$. Here, $\Delta\phi(x, y) \equiv |\phi_x - \phi_y|$ and ${\bf j}_{1,2,3}$ are the momenta of the three leading jets in the event. The Higgs boson candidate is reconstructed from the momenta of the two leading $b$--jets. Therefore, we require that all the events have exactly two $b$--jets. To be sure that we restrict the analysis in the resolved regime, we further require that $E_{T}^{\rm miss} \in~]300, 1000]~{\rm GeV}$. Note that this requirement is slightly different from the ATLAS definition of the resolved region \cite{ATLAS:2021shl}. Higgs boson candidates are required to have a transverse momentum that is larger than $300$ GeV. To ensure that reconstructed Higgs boson candidate and the missing transverse momentum (${\bf p}_{\rm miss}$) are back-to-back, we require that $\Delta\phi({\bf p}_{\rm miss}, {\bf p}_{\rm H}) > 2 \pi/3$. Backgrounds where $E_{T}^{\rm miss}$ originates from leptonically-decaying $W$--boson have the particular property that the transverse mass formed by $E_{T}^{\rm miss}$ and either the leading or the subleading $b$--jets to be bounded by the top quark mass from above. These two variables are defined as
\begin{eqnarray}
    m_{T}^{\rm min} &\equiv& \sqrt{2~p_T^{\rm slead}~E_{T}^{\rm miss} (1 - \cos\Delta\phi({\bf b}_{\rm slead}, {\bf p}_{\rm miss}))}, \nonumber \\
    m_{T}^{\rm max} &\equiv& \sqrt{2~p_T^{\rm lead}~E_{T}^{\rm miss} (1 - \cos\Delta\phi({\bf b}_{\rm lead}, {\bf p}_{\rm miss}))}, 
\end{eqnarray}
with $p_T^{\rm lead}$ and $p_T^{\rm slead}$ refer to the $p_T$ of the leading and the subleading $b$-jets respectively and $\Delta\phi(x, y) \equiv |\phi_a - \phi_b|$. We require that $m_T^{\rm min} > 170$ GeV and $m_T^{\rm max} > 200$ GeV. The total number of jets that includes both tagged and untagged jets is required to be less than 3. The effect of this cut is however very minor on all the processes. Finally, we require that the invariant of the Higgs candidate to satisfy 
$$
80~{\rm GeV} < m_{b\bar{b}} < 160~{\rm GeV}.
$$

\begin{table}[!t]
  \begin{center}
\setlength\tabcolsep{3pt}
    \begin{tabular}{l cc cc cc cc}
    \toprule \toprule
&  \multicolumn{2}{c}{$VV+X$} & \multicolumn{2}{c}{$t\bar{t}+X$} & \multicolumn{2}{c}{$H+X$} & \multicolumn{2}{c}{${\rm BP1}$} \\
 \toprule \toprule
      & Events & $\varepsilon$ & Events & $\varepsilon$ & Events & $\varepsilon$ & Events & $\varepsilon$ \\ \toprule \toprule
      Initial                                 & $7.4 \times 10^{5}$ & -  & 24367.3 & - & $5.1 \times 10^{5}$ & - & 20500.0 & -  \\
      \toprule
      $E_{T}^{\rm miss} > 100~{\rm GeV}$      & $2.5 \times 10^{5}$ $ \pm $ 90.6 & 0.332 & 17285.9 $ \pm $ 8.3 & 0.709  & $1.6 \times 10^{5}$ $ \pm $ 73.7 & 0.315  & 15153.9 $ \pm $ 9.2 & 0.739  \\
      \toprule
      Lepton veto                             & $1.6 \times 10^{5}$ $ \pm $ 67.0 & 0.670 & 13647.4 $ \pm $ 8.0 & 0.790  & $1.5 \times 10^{5}$ $ \pm $ 72.5 & 0.961  & 15108.2 $ \pm $ 9.3 & 0.997 \\
      \toprule
      $\tau$ veto                             & $1.5 \times 10^{5}$ $ \pm $ 60.1 & 0.898 & 12061.3 $ \pm $ 7.6 & 0.884  & $1.5 \times 10^{5}$ $ \pm $ 72.1 & 0.966  & 14880.0 $ \pm $ 9.3 & 0.985 \\
      \toprule
      Photon veto                             & $1.4 \times 10^{5}$ $ \pm $ 56.2 & 0.937 & 10998.0 $ \pm $ 7.2 & 0.912  & $1.5 \times 10^{5}$ $ \pm $ 72.0 & 0.990  & 14622.4 $ \pm $ 9.4 & 0.983 \\
      \toprule
      $\geq 2~{\rm small}$-${\rm R~jets}$     & $1.0 \times 10^{5}$ $ \pm $ 41.6 & 0.727 & 8967.4 $ \pm $ 6.1 & 0.815  & 82718.7 $ \pm $ 46.7 & 0.561 & 10729.8 $ \pm $ 8.9 & 0.734 \\
      \toprule
      $\geq 2~b$-jets                  & 1904.6 $ \pm $ 0.6 & 0.019 & 1920.4 $ \pm $ 1.6 & 0.214  & 19203.6 $ \pm $ 12.8 & 0.232  & 929.2 $ \pm $ 1.1 & 0.087 \\
      \toprule
      $\Delta\phi > 0.35$                     & 1732.2 $ \pm $ 0.6 & 0.909 & 1190.6 $ \pm $ 0.9 & 0.620  & 19155.7 $ \pm $ 13.1 & 0.998  & 769.3 $ \pm $ 0.9 & 0.828 \\
      \toprule
      $N_b = 2$                               & 1617.7 $ \pm $ 0.5 & 0.934 & 1133.2 $ \pm $ 0.9 & 0.952 & 19064.6 $ \pm $ 13.4 & 0.995   & 764.8 $ \pm $ 0.9 & 0.994 \\
      \toprule
      $E_{T}^{\rm miss} \in~]300, 1000]~{\rm GeV}$& 402.3 $ \pm $ 0.1 & 0.249 & 432.4 $ \pm $ 0.4 & 0.382  & 2195.4 $ \pm $ 1.3 & 0.115  & 549.6 $ \pm $ 0.6 & 0.719 \\
      \toprule
      $p_T^{b\bar{b}} > 300~{\rm GeV}$        & 230.7 $ \pm $ 0.1 & 0.574 & 319.5 $ \pm $ 0.3 & 0.739 & 1942.0 $ \pm $ 1.3 & 0.885 & 446.3 $ \pm $ 0.5 & 0.812 \\
      \toprule
      $\Delta\phi(\vec{p}_{\rm miss}, \vec{p}_{\rm H}) > 2 \pi/3$& 224.2 $ \pm $ 0.1 & 0.972 & 306.4 $ \pm $ 0.3 & 0.959 & 1941.7 $ \pm $ 1.3 & 1.000 & 317.7 $ \pm $ 0.4 & 0.712 \\
      \toprule
      $m_{T,b}^{\rm min} > 170~{\rm GeV}$     & 208.5 $ \pm $ 0.1 & 0.930 & 129.5 $ \pm $ 0.1 & 0.422 & 1928.3 $ \pm $ 1.4 & 0.993 & 314.0 $ \pm $ 0.4 & 0.988 \\
      \toprule
      $m_{T,b}^{\rm max} > 200~{\rm GeV}$     & 208.0 $ \pm $ 0.1 & 0.998 & 129.4 $ \pm $ 0.1 & 1.000 & 1927.6 $ \pm $ 1.4 & 1.000 & 313.6 $ \pm $ 0.4 & 0.999 \\
      \toprule
      $N_{\rm jets} < 3$                      & 183.3 $ \pm $ 0.1 & 0.882 & 81.7 $ \pm $ 0.1 & 0.632 & 1921.4 $ \pm $ 1.4 & 0.997   & 312.5 $ \pm $ 0.4 & 0.996 \\
      \toprule
      $m_{b\bar{b}} \in~]80, 160[~{\rm GeV}$& 7.5 $ \pm $ 0.0 & 0.041 & 3.6 $ \pm $ 0.0 & 0.045 & 1520.7 $ \pm $ 1.2 & 0.791 & 216.2 $ \pm $ 0.3 & 0.692  \\
      \bottomrule \bottomrule
    \end{tabular}
    \caption{The cutflow table for the event selection used in the resolved region for the backgrounds and an example of the signal event (BP1). For each entry we show the number of events after each selection step along with the statistical uncertainty. We also show the efficiency after each selection as defined in eq. \ref{eq:efficiency}.}
    \label{tab:cutflow:resolved}
  \end{center}
\end{table}

In Table \ref{tab:cutflow:resolved} we show the cutflow table in the resolved analysis for both the backgrounds ($VV+X$, $t\bar{t}+X$ and $H+X$) and the signal. For the signal we show only the example of BP1. In each selection we calculate the MC uncertainty that arise from the statistical limitation and the acceptance times the efficiency defined as 
\begin{eqnarray}
    \varepsilon \equiv \frac{N_i}{N_{i-1}},
    \label{eq:efficiency}
\end{eqnarray}
where $N_i$ and $N_{i-1}$ correspond to the number of events that survive the selection $i$ and $i-1$ respectively. We can see that the requirement on having at least two $b$--tagged jets kills about $82\%$ of the signal events which is not in agreement with the naive expectation of $\varepsilon \propto \epsilon_b^2 \approx 49\%$ ($\epsilon_b$ is the $b$--tagging efficiency)\footnote{This is due to the fact that the Higgs boson produced in association with DM particles is highly boosted and that its decay products can not be resolved. As an example, for a Higgs boson  decaying into $b\bar{b}$, the $\Delta R_{b\bar{b}}$ separation is roughly given by \cite{Bernreuther:2018nat} 
$$
\Delta R_{b\bar{b}} \propto \frac{M_{H_{\rm SM}}}{p_{T, H}} \times (z_b z_{\bar{b}})^{-1/2} 
$$ 
where $z_b$ and $z_{\bar{b}}$ are the momentum fractions for the bottom and the anti-bottom quarks. For the SM Higgs boson with transverse momentum in the range $[500, 1000]$ GeV decaying democratically, {\it i.e.} $z_b = z_{\bar{b}} = 1/2$ we have $\Delta R_{b\bar{b}} \approx 0.06$--$0.1$.}.  After all the selections we get an accumulated efficiency of about $0.9\%$--$1.2\%$ for the signal which is slightly dependent on the DM mass assumption while for the backgrounds we get an overall efficiency of $0.1\%$. We also calculate the significance using Asimov formula \cite{Cowan:2010js}. The results of this calculation for the four benchmark points are shown in Table \ref{tab:SS:resolved}.

\begin{table}[!t]
  \begin{center}
   \setlength\tabcolsep{12pt}
    \begin{tabular}{l c c c c}
    \toprule
    Benchmark point & BP1 & BP2 & BP3 & BP4 \\
    \toprule
    ${\cal S}$ & 5.40 & 1.61 & 0.14 & $2.62 \times 10^{-2}$ \\ 
    \toprule
    \end{tabular}
    \caption{Signal significance for the four benchmark points in the resolved regime.}
    \label{tab:SS:resolved}
    \end{center}
\end{table}

\subsection{Merged regime}

\begin{figure}[!t]
    \centering
    \includegraphics[width=0.329\linewidth]{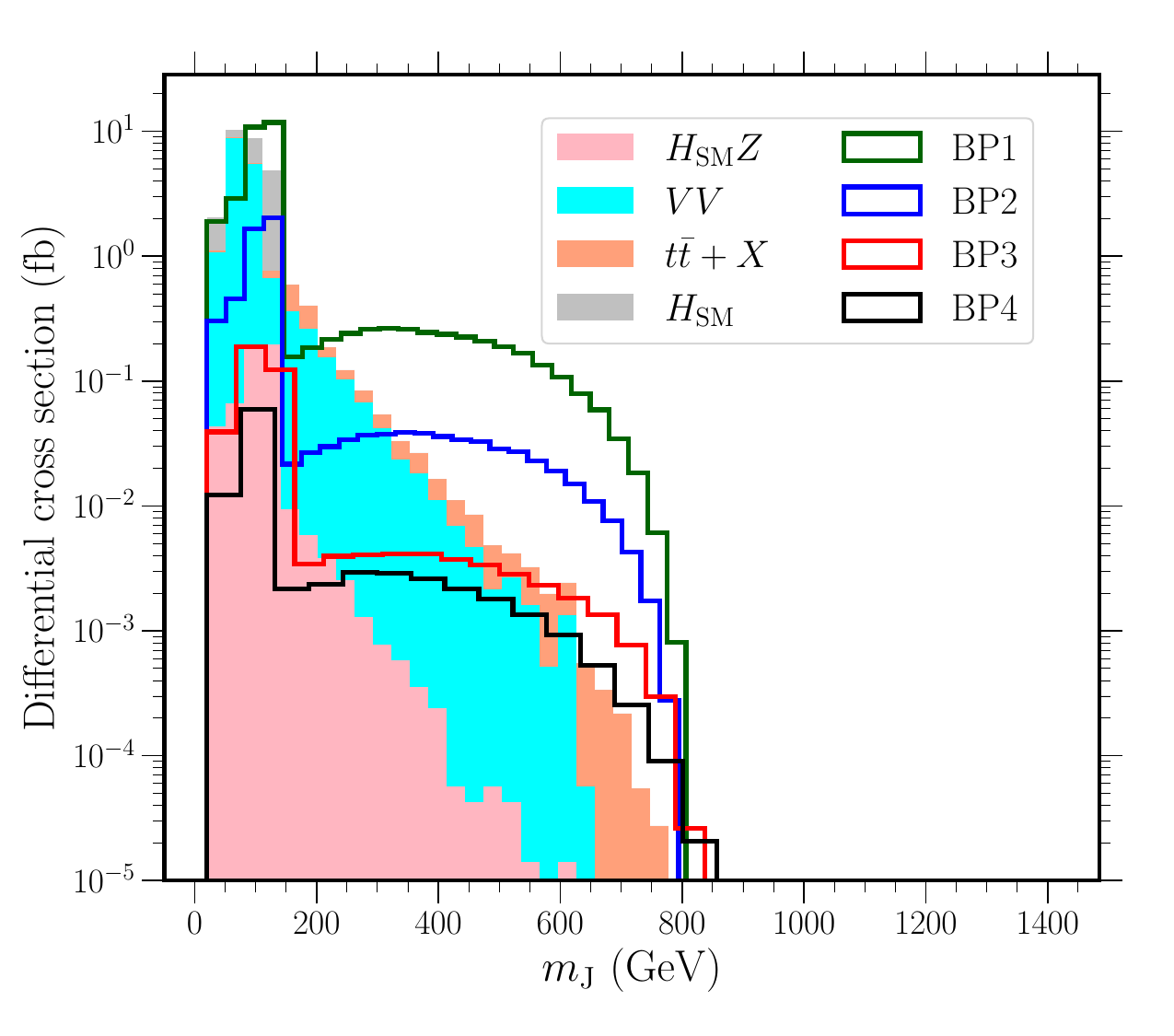}
    \hfill
    \includegraphics[width=0.329\linewidth]{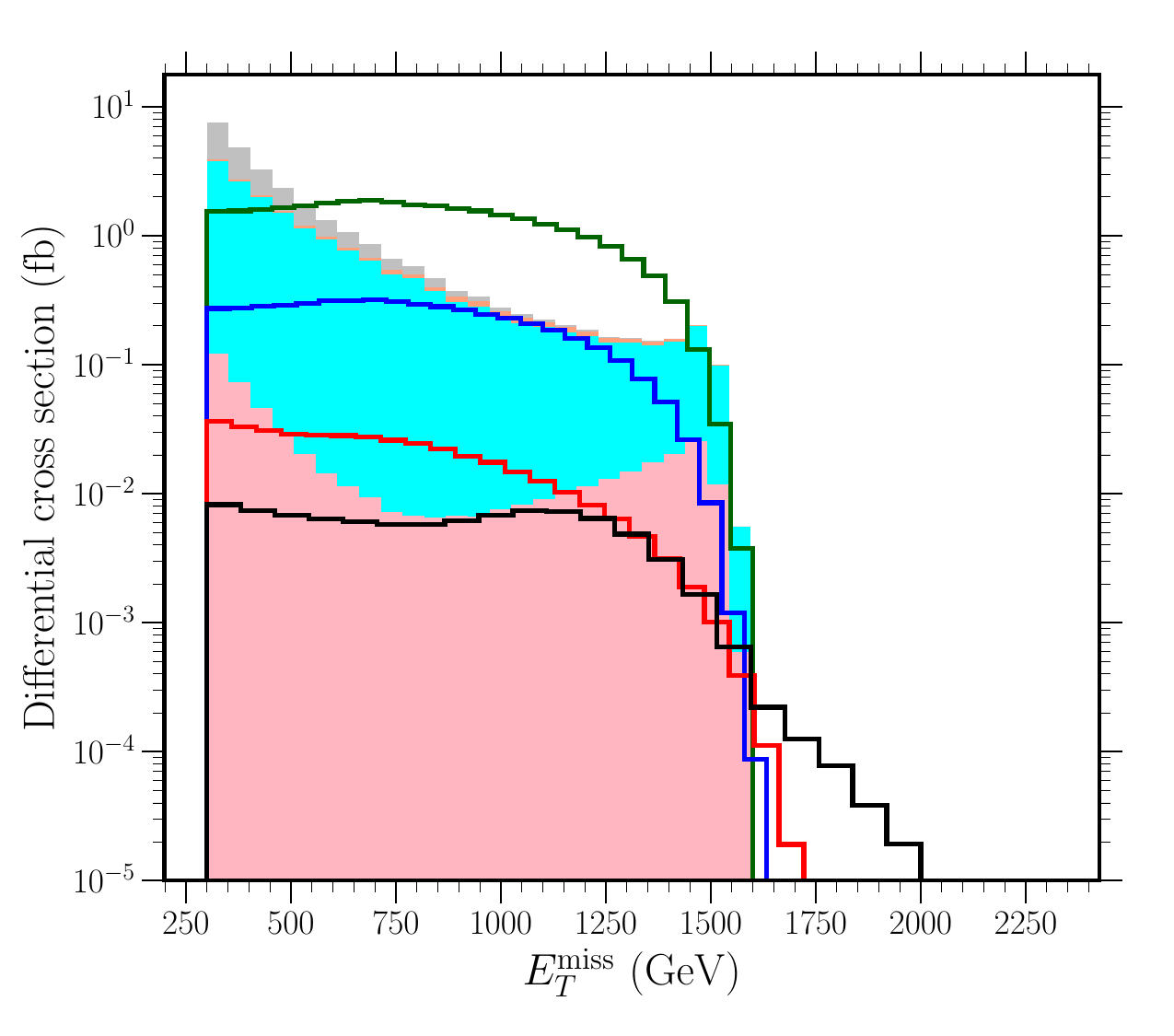}
    \hfill 
    \includegraphics[width=0.329\linewidth]{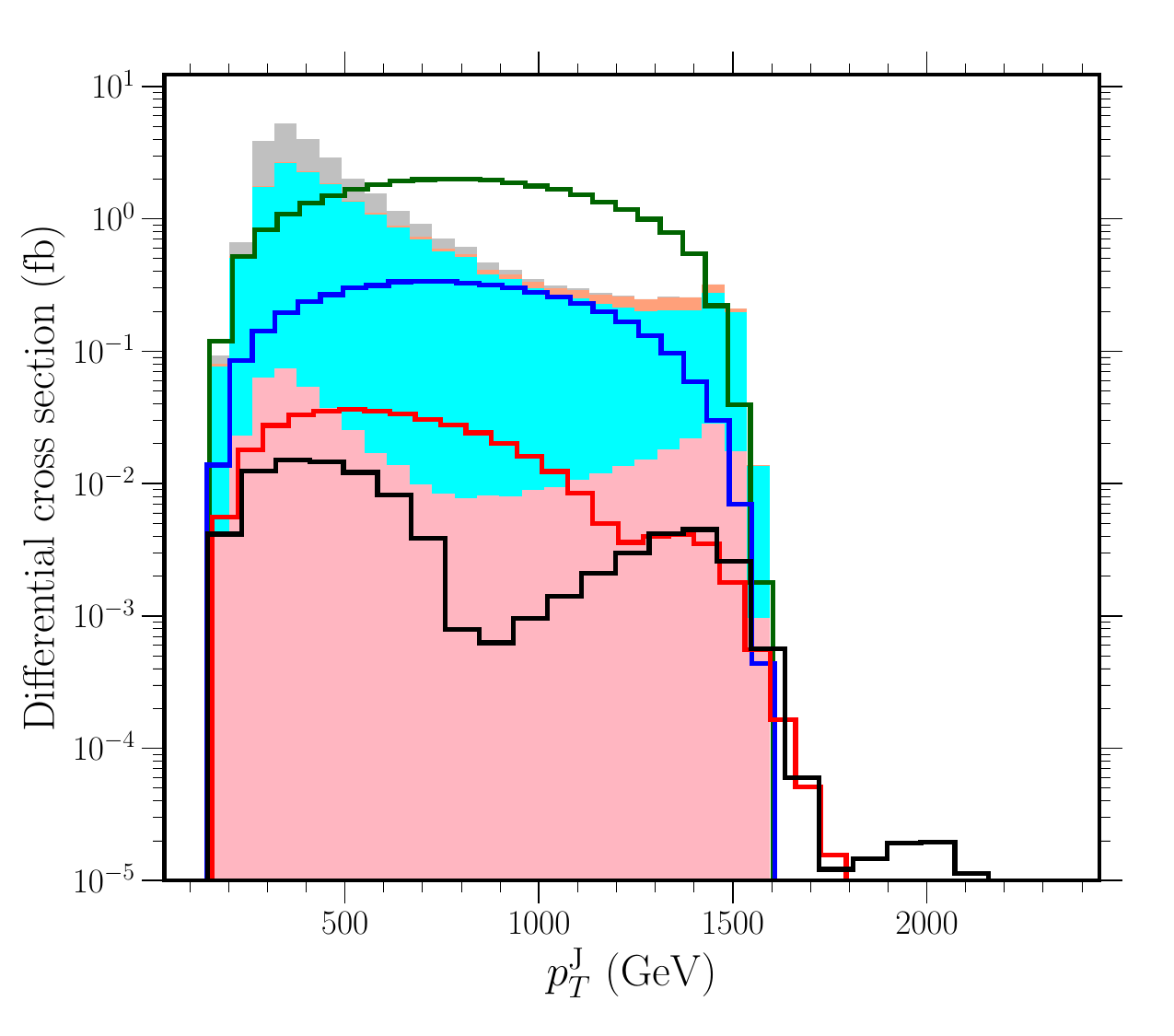}
    \caption{Differential cross section per bin for the four benchmark points defined in Table \ref{tab:BSs} and the background processes shown as stacked histograms in the merged regime for the AK10 jet category. We show the invariant mass of the leading trimmed jet $m_{\rm J}$ (left panel), the missing transverse energy $E_{T}^{\rm miss}$ (middle panel) and the transverse momentum of the leading trimmed jet $p_T^{\rm J}$ (right panel).}
    \label{fig:distributions:AK10}
\end{figure}

\begin{table}[!t]
  \begin{center}
\setlength\tabcolsep{3pt}
    \begin{tabular}{l cc cc cc cc}
    \toprule \toprule
&  \multicolumn{2}{c}{$VV+X$} & \multicolumn{2}{c}{$t\bar{t}+X$} & \multicolumn{2}{c}{$H+X$} & \multicolumn{2}{c}{${\rm BP1}$} \\
 \toprule \toprule

      & Events & $\varepsilon$ & Events & $\varepsilon$ & Events & $\varepsilon$ & Events & $\varepsilon$ \\ \toprule \toprule
      Initial                                 & $7.4 \times 10^{5}$ & -  & 24367.3 & - & $5.1 \times 10^{5}$ & - & 20500.0 & - \\
      Lepton Veto                             & $4.8 \times 10^{5}$ $ \pm $ 157.6 & 0.652 & 18123.7 $ \pm $ 7.9 & 0.744 & $4.7 \times 10^{5}$ $ \pm $ 91.4 & 0.923 & 20371.8 $ \pm $ 1.9 & 0.994 \\
      $\tau$ Veto                             & $4.6 \times 10^{5}$ $ \pm $ 156.5 & 0.956 & 16321.9 $ \pm $ 8.0 & 0.901 & $4.5 \times 10^{5}$ $ \pm $ 109.2 & 0.955 & 20029.1 $ \pm $ 3.6 & 0.983 \\
      $E_{T}^{\rm miss} > 300~{\rm GeV}$      & 39334.6 $ \pm $ 18.6 & 0.085 & 7076.6 $ \pm $ 6.1 & 0.434 & 12253.7 $ \pm $ 3.1 & 0.027 & 12014.1 $ \pm $ 9.2 & 0.600 \\
      $N_{\rm AK10}~{\rm jets} > 0$           & 37221.1 $ \pm $ 17.8 & 0.946 & 7052.4 $ \pm $ 6.0 & 0.997 & 10898.5 $ \pm $ 2.6 & 0.889 & 10923.8 $ \pm $ 8.9 & 0.909 \\
      $M_{\rm J} \in~]70,~180[~{\rm GeV}$ & 31233.4 $ \pm $ 15.0 & 0.839 & 4975.3 $ \pm $ 4.5 & 0.705 & 9202.1 $ \pm $ 2.2 & 0.844 & 8377.4 $ \pm $ 7.7 & 0.767 \\
      \bottomrule \bottomrule
    \end{tabular}
    \caption{Same as in Table \ref{tab:cutflow:resolved} but for the boosted regime with AK10 jets.}
    \label{tab:cutflow:AK10}
  \end{center}
\end{table}

As was shown in the previous subsection, the SM Higgs boson produced in the mono-Higgs channel leads to unresolved hadronic decay products. Therefore, one expects that the boosted (or merged) selection would have a higher sensitivity reach. In this subsection we perform a cut-based analysis of the mono-Higgs channel using jet substructure techniques. We perform two independent search strategies inspired by the \textsc{ATLAS} \cite{ATLAS:2021shl} and \textsc{CMS} \cite{CMS:2018zjv} boosted analyses. Therefore, we employ two different jet clustering algorithms altough using the same selection criteria (see section \ref{sec:objects} for more details):

\begin{itemize}
    \item First analysis category: We cluster jets using the anti-$k_t$ algorithm and a jet radius of $R=1$ (denoted by AK10). We use a trimming algorithm that is based on the $k_t$ algorithm and removing subjets of radius $R=0.2$ that carry less than $5\%$ of the total AK10 jet energy.
    \item Second analysis category: We cluster jets using the Cambridge-Aachen algorithm and a jet radius of $R=1.5$ (denoted by CA15). We use a soft drop algorithm to remove soft and a wide-angle radiation.
\end{itemize}

\begin{figure}[!t]
    \centering
    \includegraphics[width=0.329\linewidth]{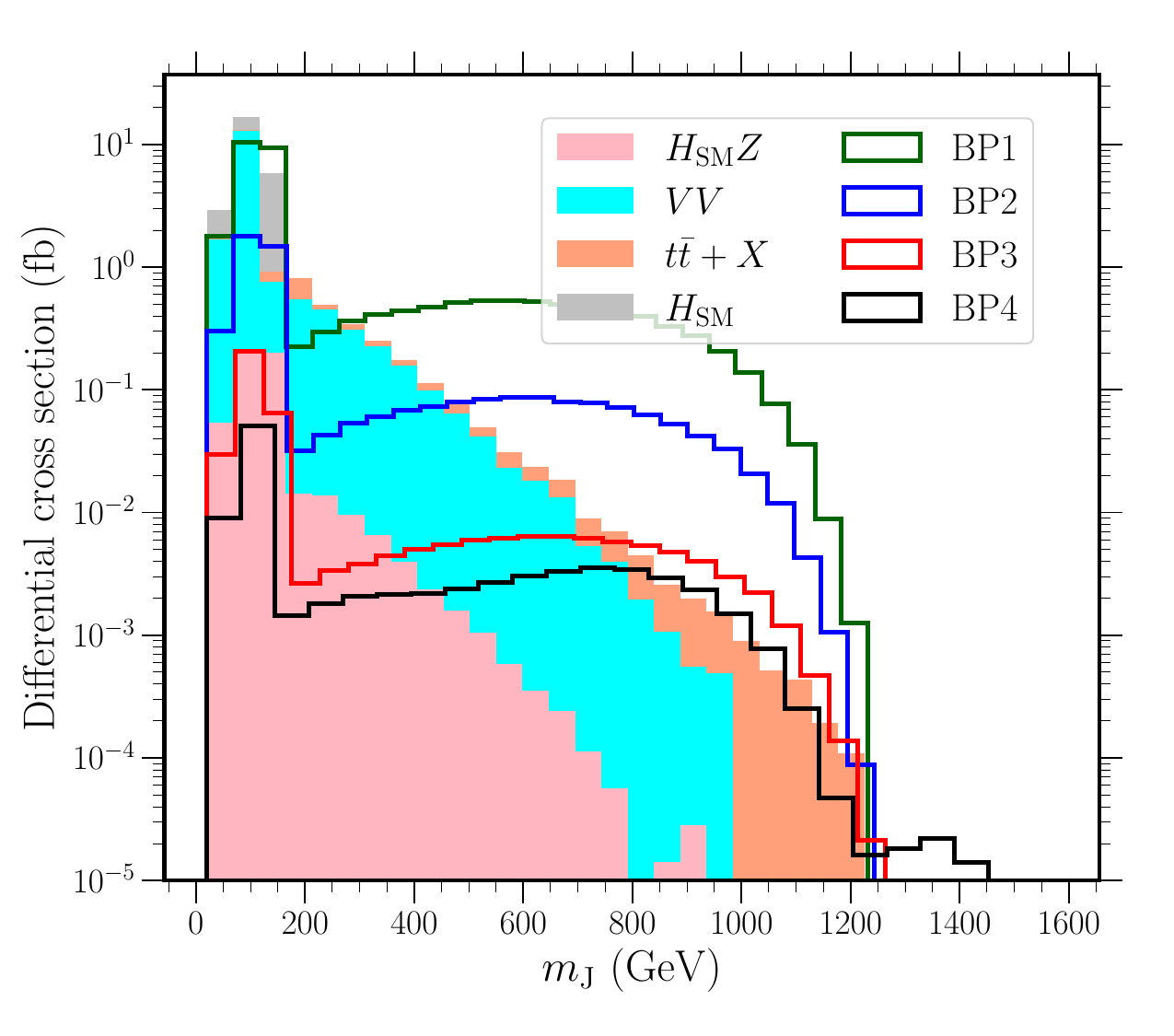}
    \hfill 
    \includegraphics[width=0.329\linewidth]{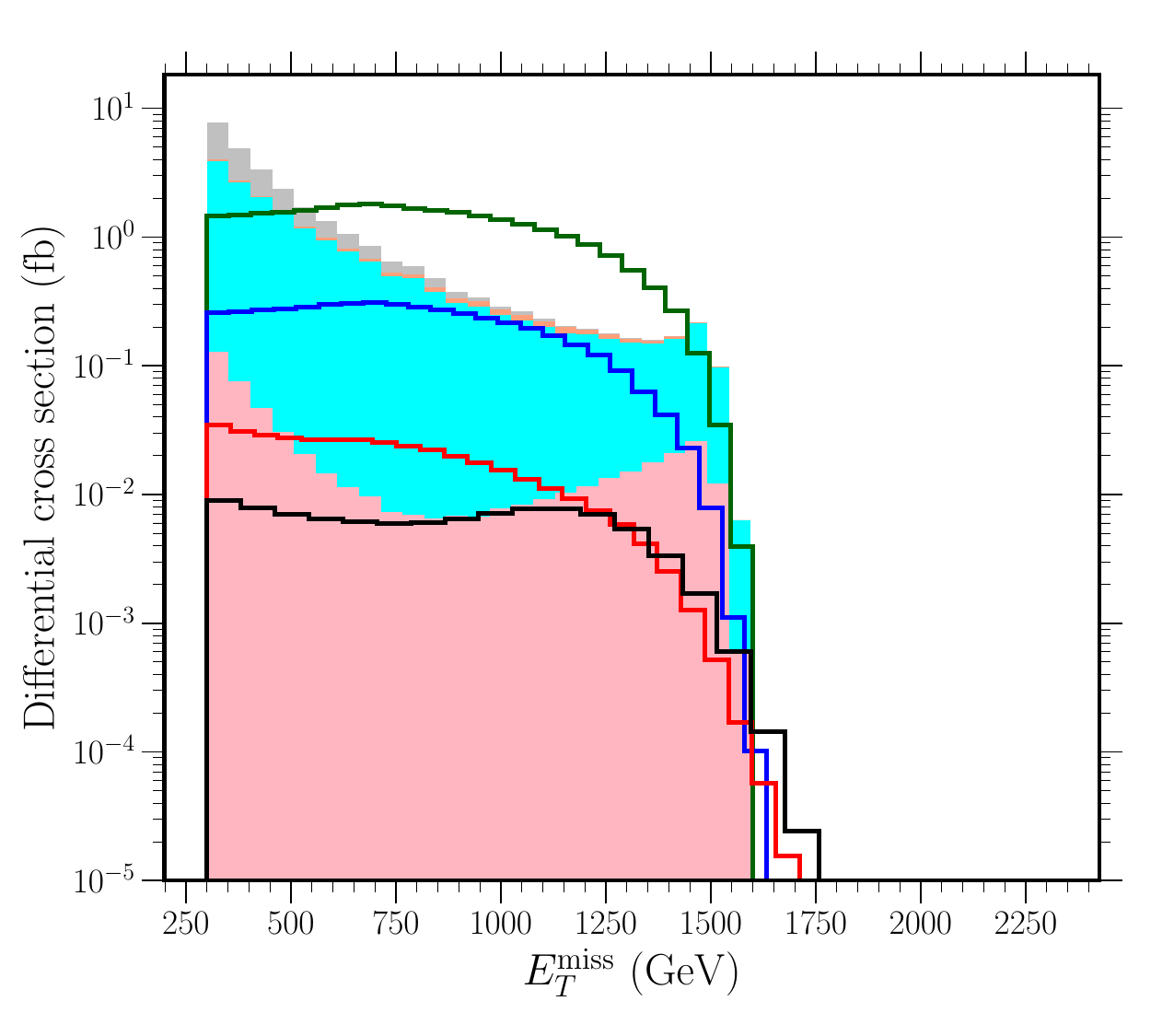}
    \hfill 
    \includegraphics[width=0.329\linewidth]{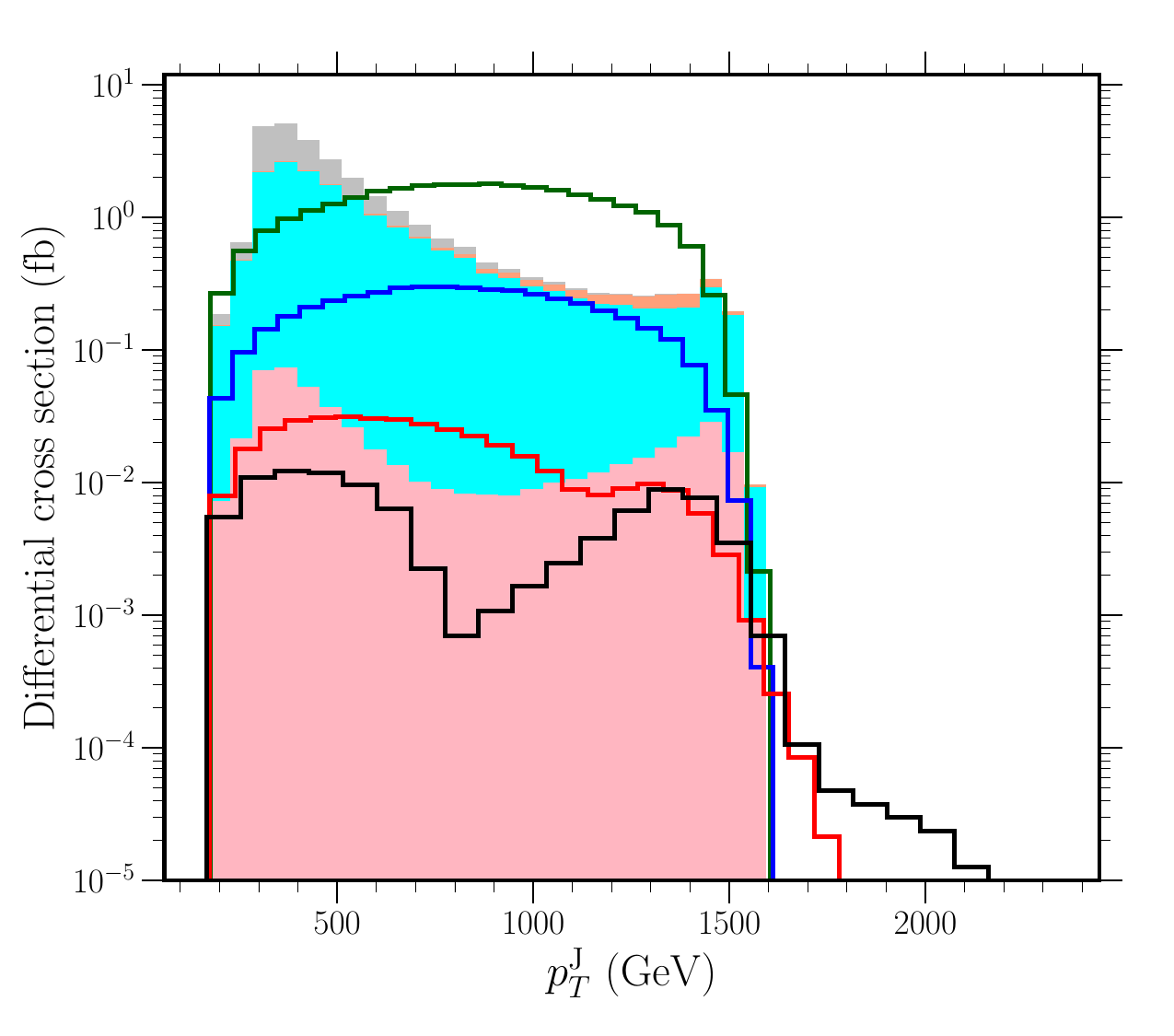}
    \caption{Differential cross section per bin for the four benchmark points defined in Table \ref{tab:BSs} and the background processes shown as stacked histograms in the merged regime for the AK10 jet category. We show the invariant mass of the leading trimmed jet $m_{\rm J}$ (left panel), the missing transverse energy $E_{T}^{\rm miss}$ (middle panel) and the transverse momentum of the leading trimmed jet $p_T^{\rm J}$ (right panel).}
    \label{fig:distributions:CA15}
\end{figure}

The differential distributions for the key observables are shown in Fig. \ref{fig:distributions:AK10} (for the selection based on AK10 jets) and in Fig. \ref{fig:distributions:CA15} (for the selection based on the CA15 jets). We require that events do not contain any isolated lepton (electron or muon) with $p_T > 7$ GeV and $|\eta| < 6$. Furthermore, we veto events that contain one hadronically decaying $\tau$--lepton having $p_T > 15$ GeV and $|\eta| < 2.5$. We then require that the missing transverse energy satisfies $E_{T}^{\rm miss} > 300$ GeV. Signal-like events are required at have at least one AK10 jet with $p_T > 150$ GeV and $|\eta| < 2.5$ or at least one CA15 jet with $p_T > 150$ GeV and $|\eta| < 2.5$. The fat jets that pass these requirements are either trimmed (for AK10 jets) or soft-dropped (for CA15 jets). Therefore one requires that events contain at least one trimmed jet for the ATLAS-like analysis or at least one soft-dropped jet for the CMS-like analysis. The leading trimmed or soft-dropped jet is required to have $M > 20$ GeV. Finally we require that the leading fat jet to have an invariant mass satisfying $70~{\rm GeV} < M < 180~{\rm GeV}$. The last cut defines our signal region.  The cutflow tables are shown in Tables \ref{tab:cutflow:AK10} and \ref{tab:cutflow:CA15}. The acceptance times the efficiency for the signal varies in the range $34\%$--$41\%$ for the AK10 jets and in the range of $29\%$--$39\%$ for the CA15 jets. Finally, we calculate both the significance (defined in eq. \ref{eq:SS:1}) and the purity\footnote{The signal purity is defined as 
$$
p = \frac{n_s}{n_s+n_b},
$$
where $n_s$~($n_b$) being the number of the signal~(background) events after the full selection.} of the signal for the four benchmark points and we display the results in Table \ref{tab:SS:boosted}. We find that the signal significance for the merged regime is a factor of $6$--$10$ larger than that in the case of the resolved regime.

\begin{table}[!t]
  \begin{center}
\setlength\tabcolsep{3pt}
    \begin{tabular}{l cc cc cc cc}
    \toprule \toprule
&  \multicolumn{2}{c}{$VV+X$} & \multicolumn{2}{c}{$t\bar{t}+X$} & \multicolumn{2}{c}{$H+X$} & \multicolumn{2}{c}{${\rm BP1}$} \\
 \toprule \toprule

      & Events & $\varepsilon$ & Events & $\varepsilon$ & Events & $\varepsilon$ & Events & $\varepsilon$ \\ \toprule \toprule
      Initial                                 & $7.4 \times 10^{5}$ & -  & 24367.3 & - & $5.1 \times 10^{5}$ & - & 20500.0 & - \\
      Lepton Veto                             & $4.8 \times 10^{5}$ $ \pm $ 157.6 & 0.652 & 18123.7 $ \pm $ 7.9 & 0.744  & $4.7 \times 10^{5}$ $ \pm $ 91.4 & 0.923 & 20371.8 $ \pm $ 1.9 & 0.994 \\
      $\tau$ Veto                             & $4.6 \times 10^{5}$ $ \pm $ 156.5 & 0.956 & 16321.9 $ \pm $ 8.0 & 0.901 & $4.5 \times 10^{5}$ $ \pm $ 109.2 & 0.955 & 20029.1 $ \pm $ 3.6 & 0.983 \\
      $E_{T}^{\rm miss} > 300~{\rm GeV}$      & 39334.6 $ \pm $ 18.6 & 0.085 & 7076.6 $ \pm $ 6.1 & 0.434 & 12253.7 $ \pm $ 3.1 & 0.027 & 12014.1 $ \pm $ 9.2 & 0.600 \\
      $N_{\rm CA15}~{\rm jets} > 0$           & 39166.6 $ \pm $ 18.6 & 0.996 & 7075.5 $ \pm $ 6.1 & 1.000 & 12182.3 $ \pm $ 3.0 & 0.994 & 12014.1 $ \pm $ 9.2 & 1.000 \\
      $N_{\rm SD}~{\rm jets} > 0$             & 37419.0 $ \pm $ 17.9 & 0.955 & 7060.5 $ \pm $ 6.0 & 0.998 & 11083.4 $ \pm $ 2.7 & 0.910 & 11118.7 $ \pm $ 9.0 & 0.925 \\
      $M_{\rm SD} \in~]70, ~180[~{\rm GeV}$ & 31615.2 $ \pm $ 15.3 & 0.845 & 4233.1 $ \pm $ 3.9 & 0.600 & 9625.0 $ \pm $ 2.3 & 0.868 & 6978.8 $ \pm $ 6.8 & 0.628 \\
      \bottomrule \bottomrule
    \end{tabular}
    \caption{Same as in Table \ref{tab:cutflow:resolved} but for the boosted regime with CA15 jets.}
    \label{tab:cutflow:CA15}
  \end{center}
\end{table}

\begin{table}[!t]
  \begin{center}
   \setlength\tabcolsep{12pt}
    \begin{tabular}{l c c c c c}
    \toprule
    Benchmark point & & BP1 & BP2 & BP3 & BP4 \\
    \toprule
    \multirow{2}{*}{${\cal S}$} & AK10 jets & $38.18$ & $10.91$ & $0.84$ & $0.17$ \\
    & CA15 jets & $31.93$ & $9.02$ & $0.74$ & $0.15$ \\
    \toprule
    \multirow{2}{*}{$p$} & AK10 jets & $0.15$ & $4.91 \times 10^{-2}$ & $3.95 \times 10^{-3}$ & $8.23 \times 10^{-4}$ \\
    & CA15 jets & $0.13$ & $4.08 \times 10^{-2}$ & $3.46 \times 10^{-3}$ & $7.06 \times 10^{-4}$ \\
    \toprule
    \end{tabular}
    \caption{Signal significance (${\cal S}$) and purity ($p$) for the four benchmark points in the boosted regime for the AK10 jets (first rows) and CA15 jets (second rows).}
    \label{tab:SS:boosted}
    \end{center}
\end{table}

\section{Optimisation using Boosted-Decision Trees}
\label{sec:BDTs}

\subsection{General setup}

\begin{figure}[!t]
    \centering
\includegraphics[width=0.329\linewidth]{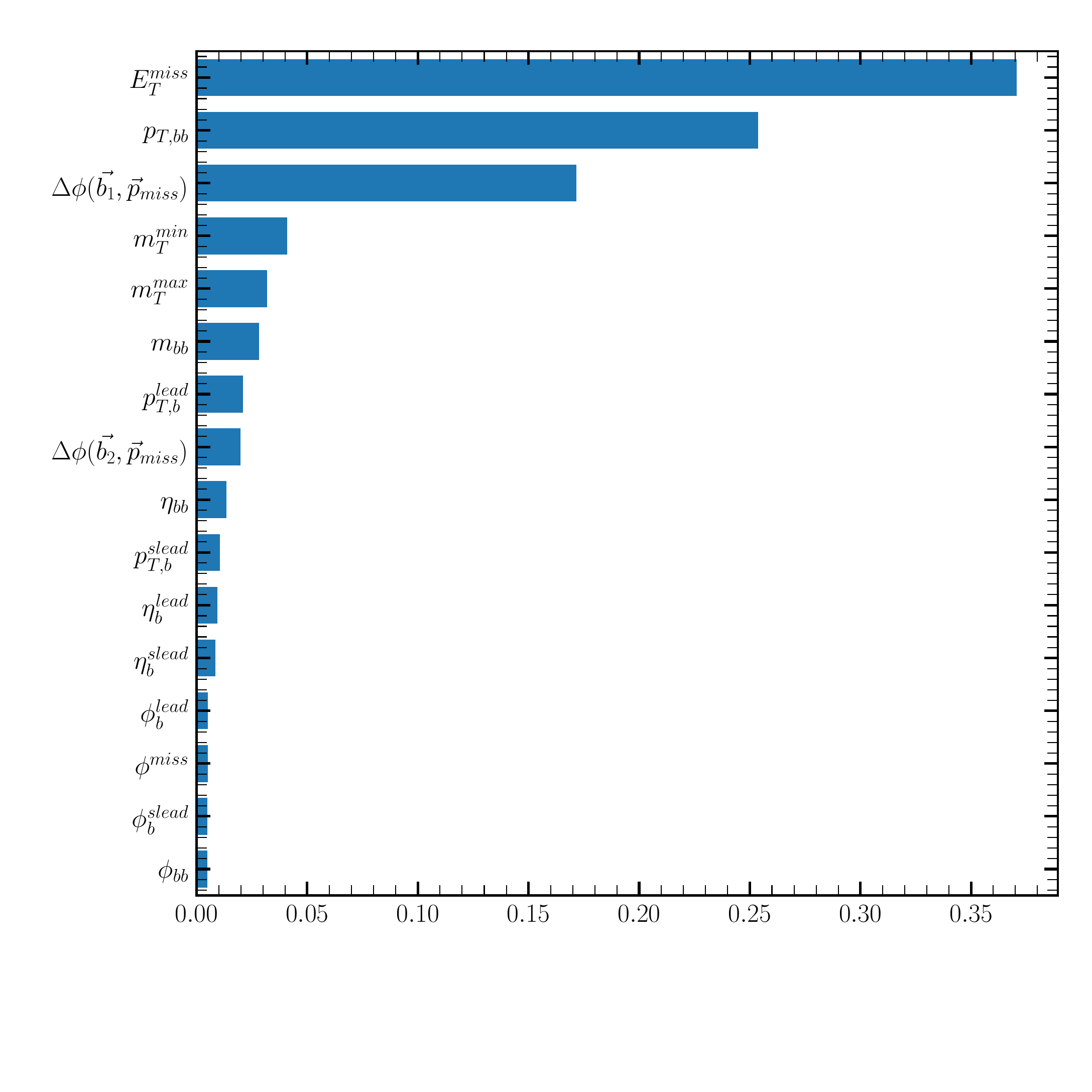}
\hfill 
\includegraphics[width=0.329\linewidth]{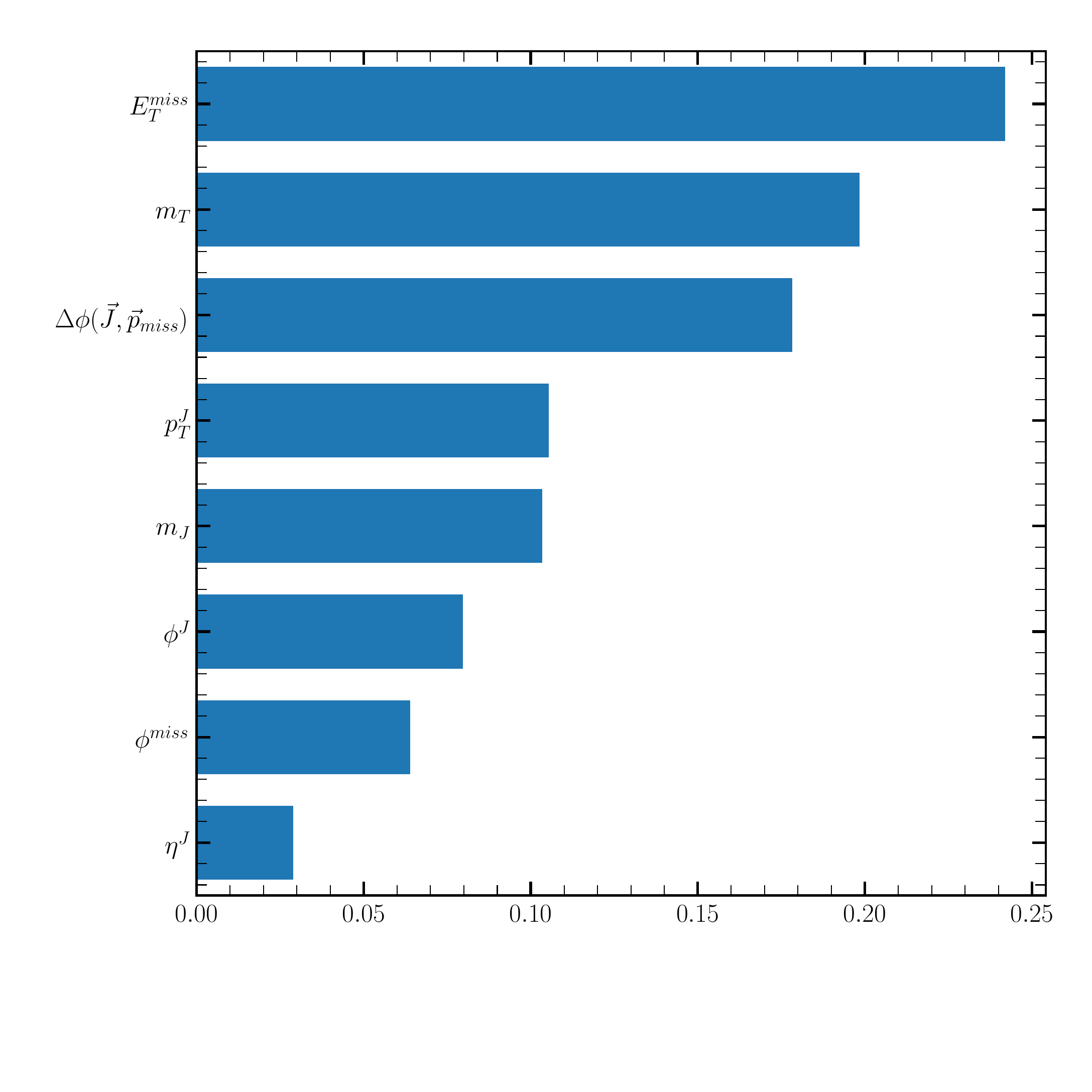}
\hfill
\includegraphics[width=0.329\linewidth]{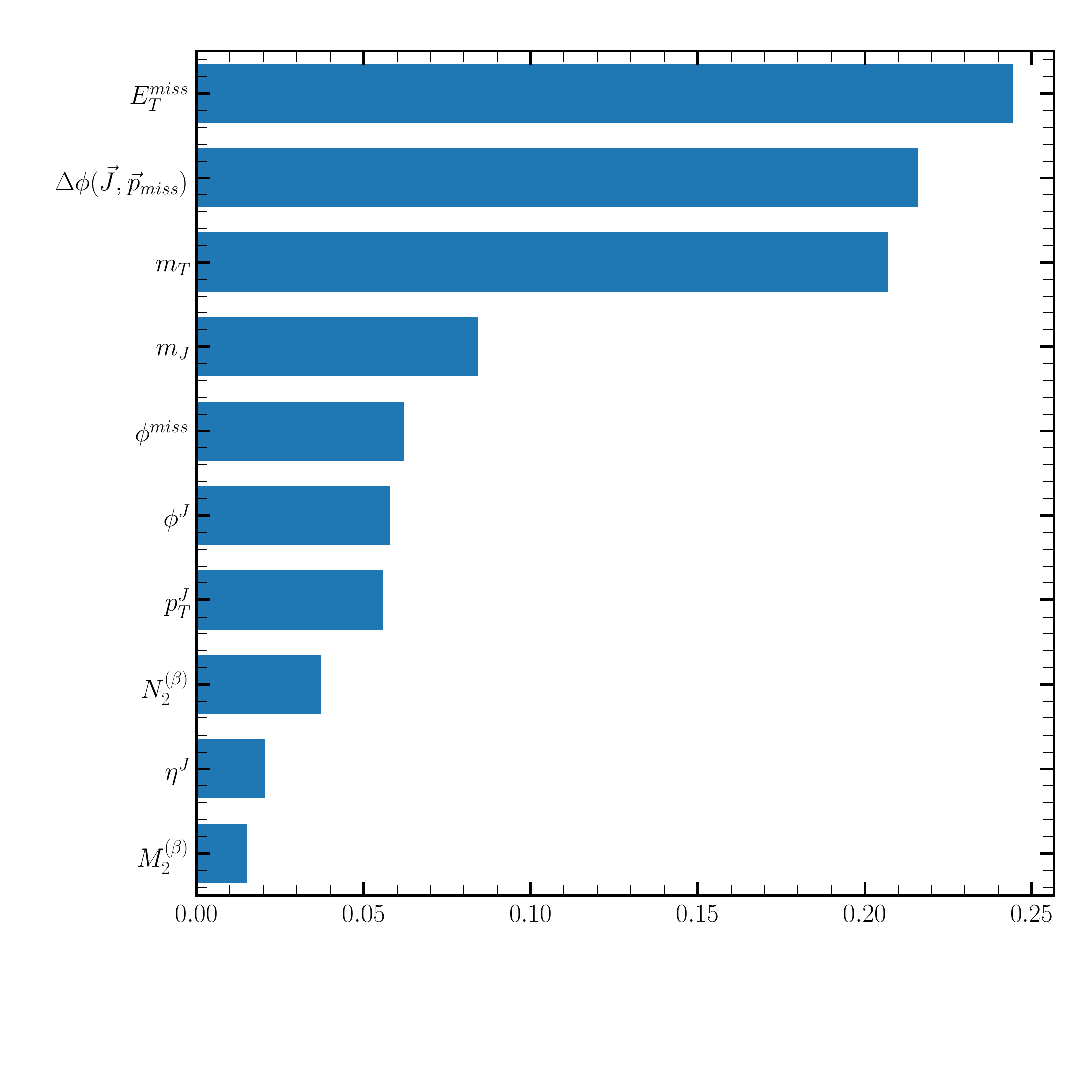}
\vspace{-1cm}
    \caption{Feature importance for the resolved regime (left panel), boosted regime with AK10 jets (middle panel) and boosted regime with CA15 jets (right panel).}
    \label{fig:feature:importance}
\end{figure}

An improvement of the previous results can be achieved by using Machine Learning (ML) algorithms, such as  decision trees (BDTs). The BDT training is performed using the four benchmark points described in Table \ref{tab:BSs} which are merged in one signal sample and for the background sample we merge all the SM background processes. All the processes are weighted by their generator-level cross sections since each process, for both the signal and the background, has a different cross section. Furthermore, in the case where the MC samples for the signal contain more events than the background samples, we reweight the signal and background samples using weight computed via the \texttt{compute\_sample\_weight} as implemented in \textsc{Scikit-Learn} \cite{scikit-learn}. The BDT algorithm was implemented using \textsc{XGBoost} classifier \cite{Chen_2016}. The model has been trained using a feature set consisting of the following variables:

\begin{itemize}
    \item {\bf Resolved regime}: 
    $$
    \{E_{T}^{\rm miss}, \phi_{\rm miss}, p_{T,b}^i, \phi_b^i, \eta_b^i, E_b^i, p_T^{bb}, \phi_{bb}, \eta_{bb}, E_{bb}, m_{bb}, \Delta\phi(\vec{b}_1, \vec{p}_{\rm miss}), \Delta\phi(\vec{b}_2, \vec{p}_{\rm miss}), m_{T}^{\rm min}, m_{T}^{\rm max}\}
    $$
    \item {\bf Boosted regime with AK10 jets}:
    $$
    \{E_{T}^{\rm miss}, \phi_{\rm miss}, p_T^{\rm J}, \eta_{\rm J}, \phi_{\rm J}, E_{\rm J}, m_{\rm J}, \Delta\phi(\vec{\rm J}, \vec{p}_{\rm miss}), m_T({\rm J}, E_{T}^{\rm miss})\}
    $$
    \item {\bf Boosted regime with CA15 jets}: 
    $$
    \{E_{T}^{\rm miss}, \phi_{\rm miss}, p_T^{\rm J}, \eta_{\rm J}, \phi_{\rm J}, E_{\rm J}, m_{\rm J}, \Delta\phi(\vec{\rm J}, \vec{p}_{\rm miss}), m_T({\rm J}, E_{T}^{\rm miss}), M_2^{(\beta)}, N_2^{(\beta)}\}
    $$
\end{itemize}
where $m_T$ is defined as 
\begin{eqnarray}
    m_T({\rm J}, E_{T}^{\rm miss}) \equiv \sqrt{2~p_T^{\rm J}~E_{T}^{\rm miss} (1 - \cos\Delta\phi(\vec{J}, \vec{p}_{\rm miss}))}
\end{eqnarray}
We briefly describe the event preselection criteria applied in our analysis. For the resolved regime, we  follow the same selection steps as in the cut-based analysis but halt  the selection process once we achieve the requirement of having exactly two $b$--tagged jets. No further cuts on the magnitude of the missing energy  are applied, except the basic requirement of $E_{T}^{\rm miss} > 100$ GeV. For the boosted regime, we do not impose requirements on the invariant mass of the trimmed leading AK10 jet or the soft-dropped CA15 jets. With these requirements, we ensure enough statistics for the training. We found that some of the variables used in this study are highly correlated to $m_{bb}$ (in the resolved regime) and to $m_{\rm J}$ (in the boosted regime). To reduce these correlations of these variables we scale $p_{T}^b$ and $p_{T}^{bb}$ by $m_{bb}$ and scale $p_T^{\rm J}$ by $m_{\rm J}$. Note that we do not apply a \texttt{StandardScaler()} function which removes the mean and reduces the variance to unity but instead, we apply a customised scaling whose aim is only for reducing the correlations. A careful inspection of the input variables through the calculation of the feature importance is very crucial to assess which of the variables can be the best signal-to-background discriminators. This can be seen in Fig. \ref{fig:feature:importance} where we show the feature importance for each of the input variables in the resolved regime (left panel), merged regime with AK10 jets (middle panel) and merged regime with CA15 jets (right panel). As expected, we can see that the missing transverse energy is the most sensitive feature for the model training. The other variables depend on the regime. For the resolved regime, the transverse momentum of the Higgs boson candidate ($p_{T, bb}$) and the azimuthal separation between the leading $b$-jet and the missing momentum are very important. For the merged regime, the transverse mass $m_T$, the azimuthal separation between the leading fat jet and the missing momentum -- $\Delta\phi(\vec{\rm J}, \vec{p}_{\rm miss})$ --, and the invariant mass of the leading fat jet ($m_{\rm J}$) are very important features.

To avoid overtraining effects, the standard procedure is to randomly split the data into two independent datasets: a training dataset and a testing dataset. Ensuring strong alignment between the trained model and the predicted testing data serves as a good indicator of the absence of overtraining effects. However, in this study, we adopt an alternative approach.  A cross-validation strategy with 5 folds is employed for the training: the data is split into 5 equal parts, a BDT model is trained on each fold and applied to the remaining 4 folds, and the final BDT score is taken to be the average of the 5 BDT model outputs. The 5 BDT models used the exact hyperparameters were optimized using the grid-search technique. The optimized hyperparameters are given in Table \ref{tab:hyperparameters}. To define the final BDT score binning, the BDT score (the average of 5 BDT model outputs) is scanned for maximum significance using the Asimov formula. Each BDT bin is required to have at least one background event to ensure good statistics. The result of the scan shows that the BDT score bin $[0.99, 1]$ gives the highest significance (as expected) for the different benchmarks, thus this bin is used to define the signal region. 

\begin{table}[!t]
  \begin{center}
   \setlength\tabcolsep{6pt}
    \begin{tabular}{l l c c}
    \toprule
    \toprule
    Parameter & Purpose & Default value & This work \\
    \toprule
    \texttt{subsample} & Subsample ratio of the training instances & 0.5 & 0.8  \\
    \toprule
    \texttt{scale\_pos\_weight} & Control the balance between positive and negative weights & 1 & 6 \\
    \toprule
    \texttt{reg\_lambda} & L2 regularization term on the weights & 0 & 10 \\
    \toprule
    \texttt{reg\_alpha} & L1 regularization term on the weights & 0 & 5 \\
    \toprule
    \texttt{n\_estimators} & The number of runs to learn from data & $-$ & 750 \\
    \toprule
    \texttt{min\_child\_weight} & Minimum sum of instance weight (hessian) needed in a child & 1 & 2 \\
    \toprule
    \texttt{max\_depth} & Maximum depth of a tree & 6 & 8 \\
    \toprule
    \texttt{learning\_rate} & Step size shrinkage used in update to prevents overtraining & 0.3 & 0.1 \\
    \toprule
    \texttt{colsample\_bytree} & The subsample ratio of columns when constructing each tree & 0.5 & 0.7 \\
    \toprule
    \texttt{tree\_method} & The tree construction algorithm used in XGBoost & 'auto' & 'hist' \\
    \toprule
    \toprule
    \end{tabular}
    \caption{Hyperparameters of the training models used in this analysis.}
    \label{tab:hyperparameters}
    \end{center}
\end{table}

\begin{figure}[!t]
    \centering
    \includegraphics[width=0.45\linewidth]{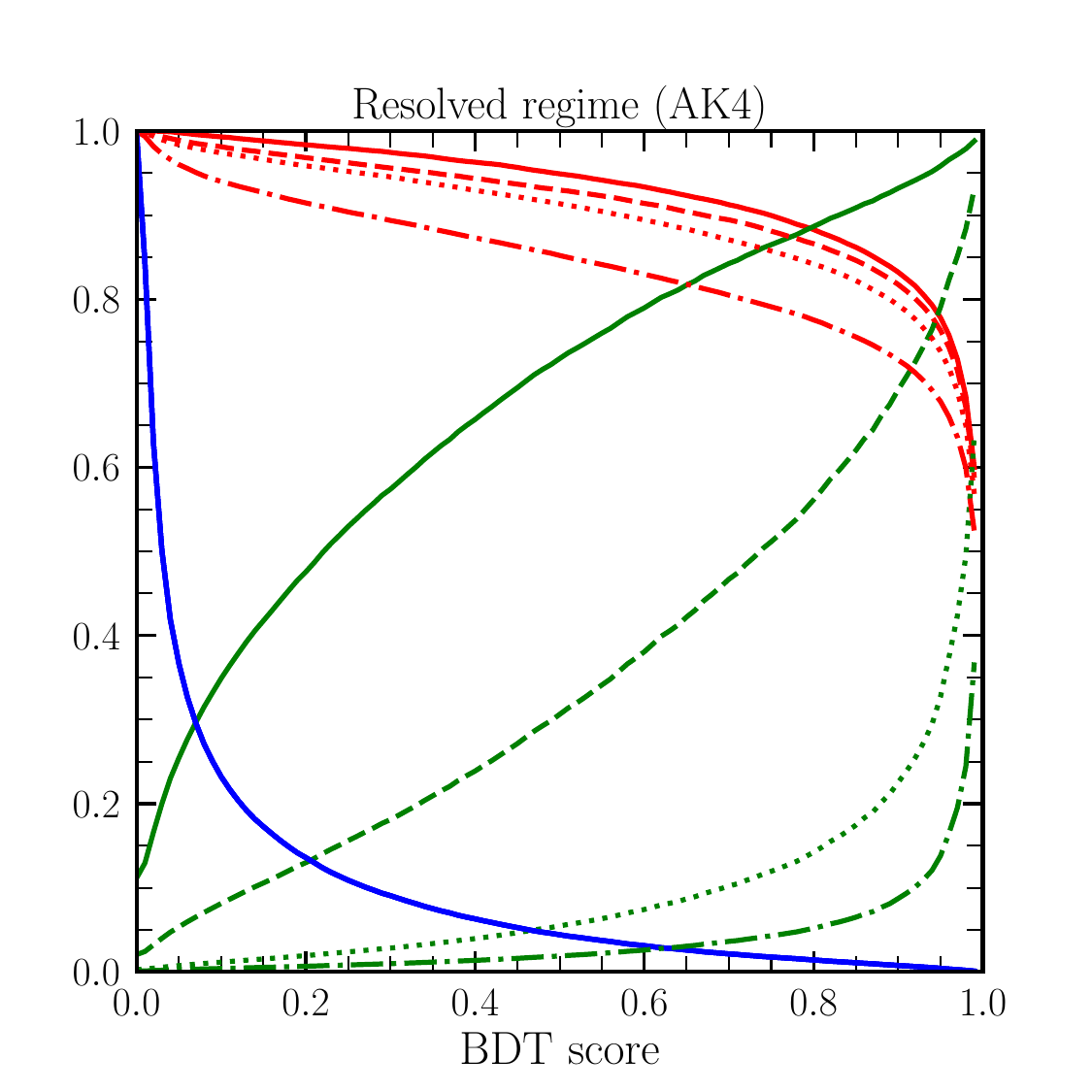}
    \hspace{-0.8cm}
    \includegraphics[width=0.45\linewidth]{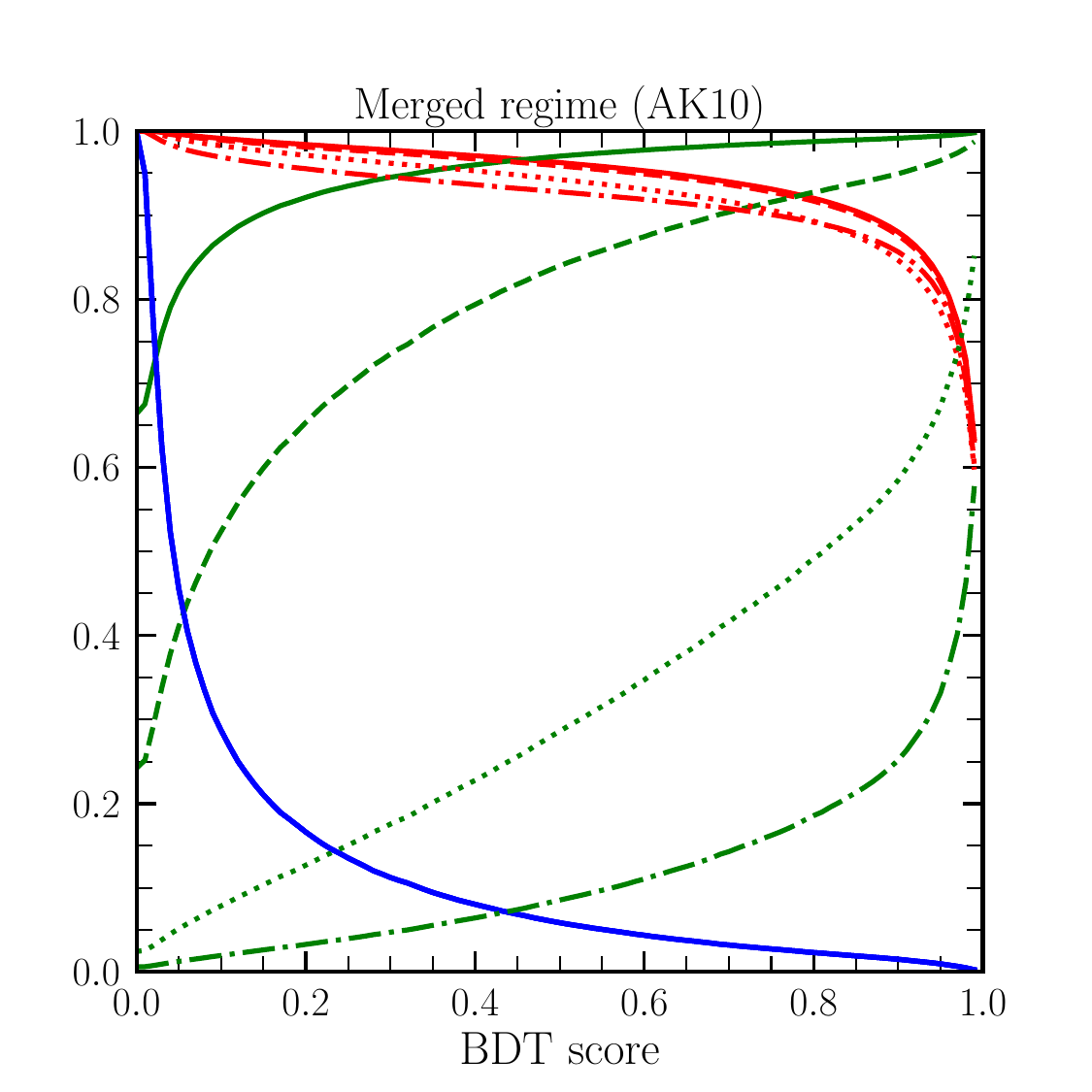}
    \vfill
    \includegraphics[width=0.51\linewidth]{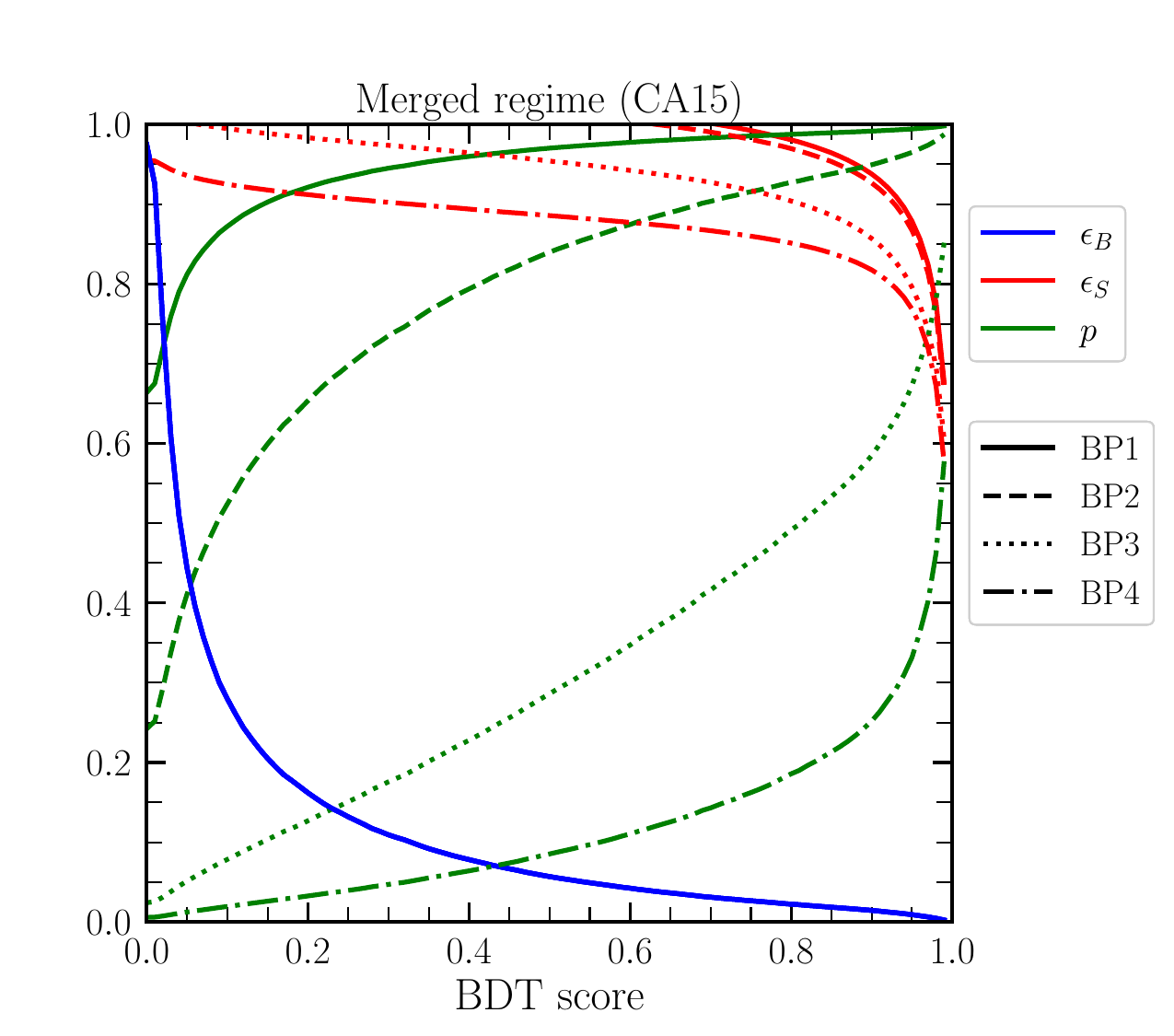}
    \caption{The background efficiency (blue), signal efficiency (red) and signal purity (green) as a function of the cut on the BDT score. Results are shown for the resolved regime (left upper panel), the merged regime with AK10 jets (right upper panel) and the merged regime with CA15 jets (lower panels). The calculations are done for BP1 (solid), BP2 (dashed), BP3 (dotted) and BP4 (dashdotted).}
    \label{fig:efficiency:purity:BPs}
\end{figure}

\subsection{Results}
\label{sec:results}

In this section we discuss the results of the BDT analysis. We first start by showing the signal purity ($p$), background efficiency~($\epsilon_B$), and the signal efficiency~($\epsilon_S$) as a function of the BDT score for the four benchmark points in figure \ref{fig:efficiency:purity:BPs}. We can see that the bin with the highest BDT score ($ > 0.99$) does not only maximise the significance but also the signal purity. The number of events for the signal is found to be quite large for most of the benchmark points with the results being more important for the merged regime than in the resolved regime. The signal purity varies in the range of $40\%$--$99\%$ where higher numbers are reached for the benchmark points BP1 and BP2. High values of the signal purity implies unprecedented opportunities to perform post-discovery analyses to assess the nature of DM at muon colliders. We must stress that our analysis has a very high accuracy since the area under the Receiver-Operating Characteristic (ROC) curve varies in the range of $0.95$--$0.97$. \\ 

We also calculate the signal significance for the signal using Asimov formula \cite{Cowan:2010js}
\begin{eqnarray}
    {\cal S} \equiv \sqrt{2 \bigg((n_s + n_b) \log\bigg(1 + \frac{n_s}{n_b}\bigg) - n_s\bigg)},
    \label{eq:SS:1}
\end{eqnarray}
for both ${\cal L} = 100~{\rm fb}^{-1}$ and ${\cal L} = 1~{\rm ab}^{-1}$ for the BDT bin $> 0.99$. The results are shown in Table \ref{tab:SS:BPs:BDT}. We can see that even for a luminosity of $100$ fb$^{-1}$ the BDT search strategy leads to quite large signal significance for BP1, BP2 and BP3 where high values are reached for the boosted regime as expected. To reach a high signal significance for BP4 (corresponding to heavy DM), the full luminosity of $1$ ab$^{-1}$ is required. We notice that very important improvements with respect to the results of the cut-based analysis are reached when comparing the results of Table \ref{tab:SS:BPs:BDT} with those shown in Tables \ref{tab:SS:resolved} and \ref{tab:SS:boosted}. The results are improved by about $8$--$50$ depending on the benchmark point and the kinematic regime. For instance, BP4 receives the highest improvement especially for the resolved regime where ${\cal S}$ increases from $2.62 \times 10^{-2}$ to $1.42$. 

\begin{table}[!t]
  \begin{center}
   \setlength\tabcolsep{18pt}
    \begin{tabular}{l c c c c c}
    \toprule
    Benchmark point & & BP1 & BP2 & BP3 & BP4 \\
    \toprule
    Resolved (AK4) & ${\cal S}_{100~{\rm fb}^{-1}}$ & 33.85 & 9.59 & 1.77 & 0.63 \\ 
    & ${\cal S}_{1000~{\rm fb}^{-1}}$ &  75.69 & 21.45 & 3.97 & 1.42  \\
    \toprule
    Merged (AK10) & ${\cal S}_{100~{\rm fb}^{-1}}$ & 143.44 & 45.41 & 6.86 & 1.99 \\ 
    &  ${\cal S}_{1000~{\rm fb}^{-1}}$ & 320.76 & 101.55 &  15.34 & 4.45 \\
    \toprule
    Merged (CA15) & ${\cal S}_{100~{\rm fb}^{-1}}$ & 149.20 & 47.83 & 7.66 & 2.60 \\ 
    & ${\cal S}_{1000~{\rm fb}^{-1}}$ & 333.62 & 106.95 & 17.13 & 5.81 \\
    \toprule
    \end{tabular}
    \caption{Signal significance (${\cal S}$) the four benchmark points using the BDT signal region. For each entry, we show the significance for ${\cal L}=100$ fb$^{-1}$ and after the full run at ${\cal L} = 1000$ fb$^{-1}$. The results are shown for the resolved regime with AK4 jets and for the two cases of the merged regime for AK10 jets and CA15 jets.}
    \label{tab:SS:BPs:BDT}
    \end{center}
\end{table}

\begin{figure}[!t]
    \centering
    \includegraphics[width=0.49\linewidth]{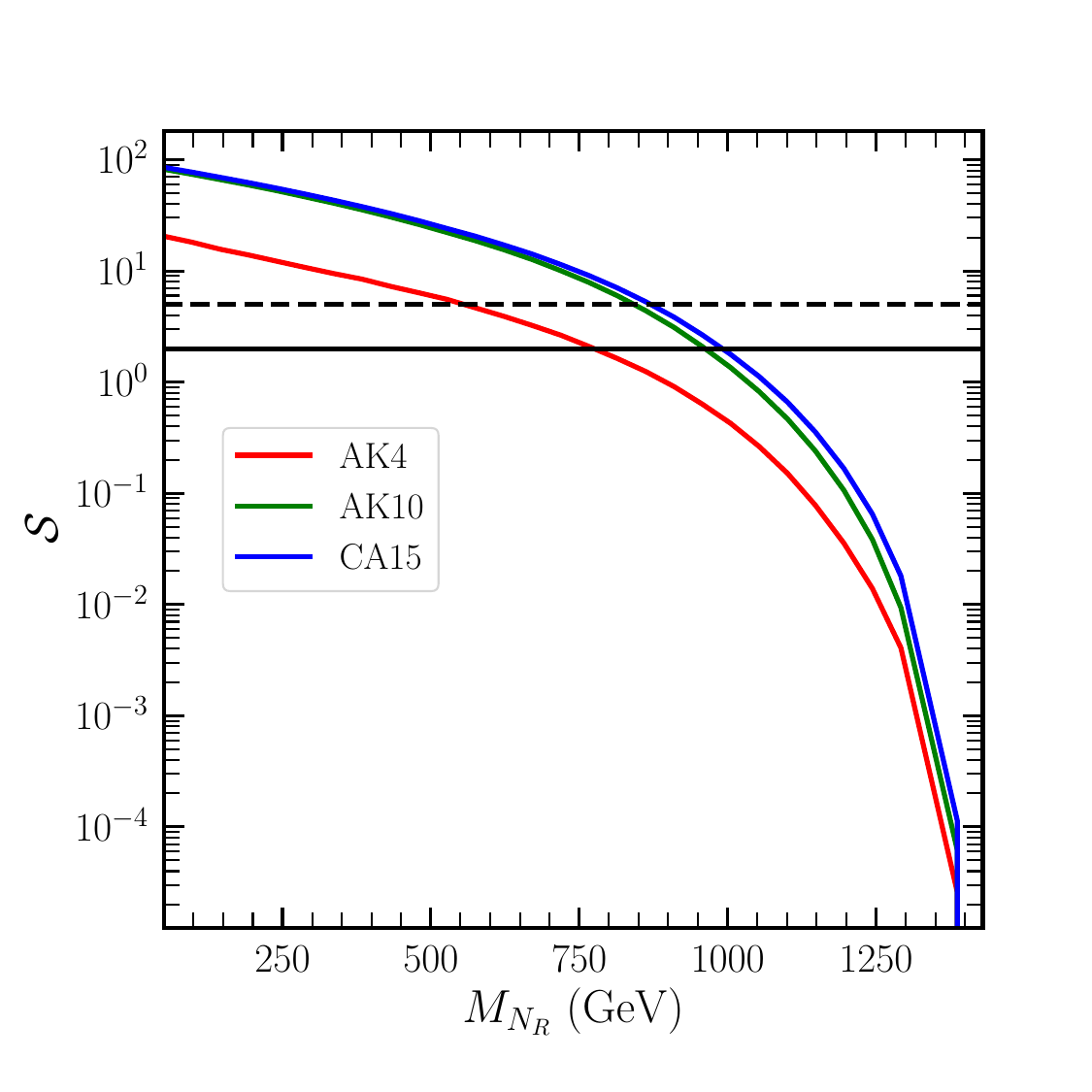}
    \hfill
    \includegraphics[width=0.49\linewidth]{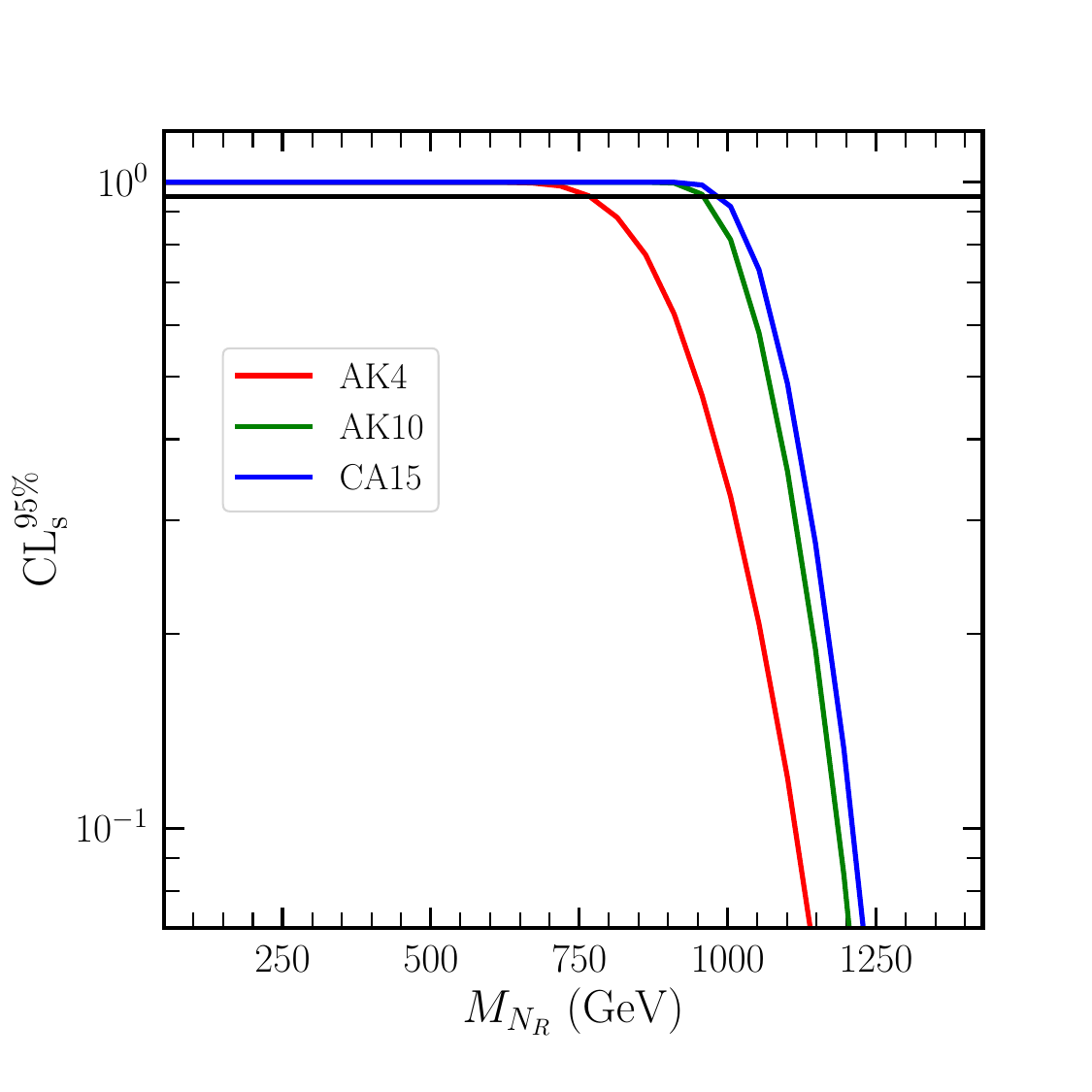}
    \caption{Signal significance (left) and ${\rm CL}_s^{95\%}$ (right) as a function of the DM mass ($M_{N_R}$). The results are shown for the resolved regime (red), boosted regime with AK10 jets (green) and the boosted regime with CA15 jets (blue). The black solid and dashed lines in the left panel correspond to ${\cal S} = 2$ and ${\cal S} = 5$. On the right panel, the solid black line corresponds to ${\rm CL}_s = 0.95$ above which the mass value is excluded at $95\%$ CL.}
    \label{fig:SS:versus:M}
\end{figure}

Finally, we use the trained algorithm to optimise the signal-over-background ratio for DM mass in the interval defined in equation \ref{eq:scan}. In other words, no further training has been performed at this stage. To quantify the sensitivity reach of this analysis we calculate both the significance and the CL$_{s}^{95\%}$. The significance is calculated by assuming some uncertainties on the background yields and is defined as 

\begin{eqnarray}
{\cal S} &=&\sqrt{2}\left[(n_s+n_b)\log\left(\frac{(n_s+n_b)(n_b+\delta_b^2)}{n_b^2+(n_s+n_b)\delta_b^2}\right) - \frac{n_b^2}{\delta_b^2} \log\left(1+\frac{\delta_b^2 n_s}{n_b(n_b+\delta_b^2)}\right)\right]^{1/2},
\label{eq:SS:2}
\end{eqnarray}
where $\delta_b = x \times n_b$ is the uncertainty on the background yields which is assumed to $x = 5\%$. Moreover we assess the sensitivity reach by computing the expected ${\rm CL}_s$ \cite{Read:2002hq} using \texttt{Pyhf} \cite{pyhf_joss}. The CL$_s$ estimator is given by
\begin{eqnarray}
    {\rm CL}_s \equiv \max\bigg(0, 1-\frac{p_{b+s}}{p_b}\bigg), 
\end{eqnarray}
where $p_{b+s}$ and $p_b$ are the signal-plus-background and the background probabilities respectively. In the calculation of the CL$_s$ we assume that the expected number of observed events is equal to the background expectations. Furthermore, we assume that the uncertainty on the background yield is $5\%$. The results are shown in Fig. \ref{fig:SS:versus:M} where we show the signal significance (left) and CL$_s$ (right) as a function of the DM mass for the resolved and the boosted regimes. We can see that the BDT analysis can probe DM masses up to $1$ TeV where both the two statistical prespcriptions lead to similar results. We finally the boosted regime has higher sensitivity than the resolved regime as expected. 

\section{Conclusions}
\label{sec:conclusions}

In this work we have studied the potential discovery of DM at muon colliders in the mono-Higgs channel. This production channel is very unique in the sense that it would allow the studies of the characteristics of the interactions between the mediator and the SM Higgs sector. As a proof-of-principle we have analysed this channel for the minimal lepton portal DM model which extends the SM with two $SU(2)_L$ singlets: a charged scalar that plays the role of the mediator and a right-handed fermion that assumed to be the DM candidate of the model. After studying the characterstic of the benchmark points allowed by the various constraints, we have studied the production of DM in this channel as well as all the possible backgrounds for center-of-mass energies of $3$, $10$ and $30$ TeV. We have found that the initial signal-to-background ratio for this channel, before any cuts, degrades very quickly with the center-of-mass energy. Therefore, we have analysed the sensitivity reach only for $\sqrt{s}=3$ TeV and ${\cal L} = 1$ ab$^{-1}$. We then performed simple cut-based analysis strategies inspired by the previous ATLAS and CMS searches of DM produced in association with a Higgs boson decaying into bottom quarks. Using different jet clustering algorithms to reconstruct the Higgs boson candidates, we have found poor sensitivities for benchmark points corresponding to heavy DM masses. We have then built an algorithm based on Boosted-Decision Trees (BDT) using \textsc{XGBoost} library. By optimising the hyperparameters of the models and training it on both the signal and the backgrounds we have found very good improvements by factors of $8$--$50$ with respect to the cut-based analysis. Finally, we have analysed the sensitivity reach by applying this algorithm to the range of DM masses kinematically allowed by the used center-of-mass energy, {\it i.e.} $M_{N_R} \in~[50, 1435]~{\rm GeV}$. We have found that DM masses up to $1$ TeV can be excluded at the $95\%$ CL using the BDT analysis and the mono--Higgs channel.

\section*{Acknowledgements}
A.J. would like to thank Jack Araz, Benjamin Fuks and Richard Ruiz for the useful discussions. The work of A.J. is supported by the Institute for Basic Science (IBS) under the project code, IBS-R018-D1.  The work of S.N. is supported by the United Arab Emirates University (UAEU) under UPAR Grant No. 12S093.

\bibliographystyle{utphys}
\bibliography{main.bib}

\end{document}